\documentclass[british,ss]{imsart}
\usepackage[T1]{fontenc}
\usepackage[latin9]{inputenc}
\usepackage{geometry}
\usepackage{babel}
\usepackage{float}
\usepackage{wrapfig}
\usepackage{textcomp}
\usepackage{amsmath}
\usepackage{amssymb}
\usepackage{graphicx}
\usepackage[authoryear]{natbib}
\usepackage[unicode=true]
 {hyperref}

\makeatletter

\providecommand{\tabularnewline}{\\}
\floatstyle{ruled}
\newfloat{algorithm}{tbp}{loa}
\providecommand{\algorithmname}{Algorithm}
\floatname{algorithm}{\protect\algorithmname}

\makeatother

\usepackage{listings}

\begin{document}
\begin{frontmatter}
\title{Nested Sampling Methods}
\runtitle{Nested Sampling Methods}
\runauthor{Johannes Buchner}
\begin{aug}
\author[A,B,C,D]{\fnms{Johannes} \snm{Buchner} \ead[label=e1]{johannes.buchner.acad@gmx.com}}
\address[A]{Max Planck Institute for Extraterrestrial Physics, Giessenbachstrasse, 85741 Garching, Germany, \printead{e1}}
\address[B]{Millenium Institute of Astrophysics, Vicu\~{n}a. MacKenna 4860, 7820436 Macul, Santiago, Chile}
\address[C]{Pontificia Universidad Católica de Chile, Instituto de Astrofísica, Casilla 306, Santiago 22, Chile}
\address[D]{Excellence Cluster Universe, Boltzmannstr. 2, D-85748, Garching, Germany}
\end{aug}
\begin{abstract}
Nested sampling (NS) computes parameter posterior distributions and
makes Bayesian model comparison computationally feasible. Its strengths
are the unsupervised navigation of complex, potentially multi-modal
posteriors until a well-defined termination point. A systematic literature
review of nested sampling algorithms and variants is presented. We
focus on complete algorithms, including solutions to likelihood-restricted
prior sampling, parallelisation, termination and diagnostics. The
relation between number of live points, dimensionality and computational
cost is studied for two complete algorithms. A new formulation of
NS is presented, which casts the parameter space exploration as a
search on a tree data structure. Previously published ways of obtaining
robust error estimates and dynamic variations of the number of live
points are presented as special cases of this formulation. A new online
diagnostic test is presented based on previous insertion rank order
work. The survey of nested sampling methods concludes with outlooks
for future research.
\end{abstract}

\end{frontmatter} 

\tableofcontents{}

\section{Context}

Nested Sampling (NS, \citealp{Skilling2004}) is a Monte Carlo algorithm
for computing an integral over a model parameter space. In the context
of Bayesian inference of analysing some data $D$, the integrand is
the likelihood function $L(D|\theta)$, which is marginalised over
the parameters $\theta$ according to the prior probability density
$\pi(\theta)\,d\theta$, which gives a measure of the parameter space.
Integrals over the posterior density $L(D|\theta)\times\pi(\theta)d\theta$
allow insightful statements about what model parameters regions are
probable or improbable. The integral over the entire parameter space,
$Z=\int L(D|\theta)\,\pi(\theta)d\theta$, is known as the marginal
likelihood or Bayesian evidence. The Bayes factor is the ratio of
marginalised likelihoods of two different models, $Z_{M1}/Z_{M2}$.
Multiplied by the model prior odds, the resulting posterior odds can
be interpreted as the relative evidence among these two models, given
the observed data. Selecting models using the Bayes factor can be
performed based on empirical scales \citep[e.g.,][]{jeffreys1961theory,Kass1995}
or calibrated on false decision rates \citep[e.g.,][]{Veitch2008,Buchner2019c}.
The computation of marginal likelihoods is thus generally important
for science with parametric probabilistic models \citep{evans2007discussion}.
In practice, the computation of posteriors and the integral is achieved
with Monte Carlo algorithms.

Exploring, navigating and integrating these parameter spaces can exhibit
the following challenges (classification in priv. comm. with F. Beaujeau):
\begin{itemize}
\item (\textbf{P}) \emph{Peculiar shapes} such as non-ellipsoidal (e.g.,
from non-Gaussian profiles) and non-convex posterior contours (e.g.,
bananas).
\item (\textbf{M}) \emph{Multiple, well-separated modes}, when several peaks
in the posterior with comparable probability exist. One can define
these by contours forming non-connected sets.
\item (\textbf{D}) \emph{High dimensionality} (here: intermediate: $d\sim10$,
high: $d>30$). High-dimensional spaces incur the curse of dimensionality
and geometric intuition breaks down.
\item (\textbf{I}) \emph{High information gain} makes the posterior a very
small portion (e.g., $e^{-10000}$) of the prior volume, that needs
to be identified. A useful measure of how much the likelihood updates
the prior is the information gain: $H=\int\log\frac{\pi(\theta)}{L(D|\theta)\times\pi(\theta)}\pi\,d\theta$.
\item (\textbf{T}) \emph{Phase transitions} are surprising and abrupt changes
of the accessible parameter space $X$ with increasing likelihood
$L$, i.e., when $g=\frac{d\log L}{d\log X}$ is not an up-concave
function \citep{skilling2006nested}. A illustrating example is the
spike-and-slab likelihood, where a Gaussian is summed with another,
much wider, co-centred Gaussian ($\sigma_{2}^{2}\gg\sigma_{1}^{2}$).
Between the centre and scales of order $\sigma_{1}$, $g$ decreases
quadratically approximately as $-(x-\mu)^{2}/\sigma_{1}^{2}$, but
then slows its decrease and becomes almost constant between $\sigma_{1}$
and $\sigma_{2}$, where the ``spike'' ($\sigma_{1}$) occupies
a tiny region on top of the wide slab ($\sigma_{2}$). Therefore,
the parameter space accessible at a given $L$ increases first very
slowly with decreasing $\log L$, then extremely rapidly, analogous
to the volume expansion of water as it is heated to water vapor \citep{skilling2006nested}.
Such phase transitions commonly occur when a subdominant model component
becomes relevant after a dominant component is constrained, for example
in mixture models. Likelihood plateaus can be considered extreme phase
transitions.
\end{itemize}
Naturally, a given problem can exhibit any combination of these challenges. 

Nested sampling (introduced in §\ref{sec:Introduction}) addresses
these challenges. It makes computing $Z$ practical for a wide variety
of problems. Posterior samples are simultaneously computed by NS.
Beyond the application to Bayesian inference, NS has been applied
as a general integration algorithm \citep[e.g.,][]{murray2006nested,partay2010efficient,Malakar2011,Goggans2014,Birge2012}
and to compute entropies \citep{Malakar2011,Brewer2017}.

\section{Review methodology}

This review presents NS methods developed over the last 15 years.
We conducted a systematic literature review to find works on nested
sampling. We used four sources: (1) Google Scholar was searched with
the terms ``nested sampling'' (including quotes) in Sep 2017. Of
the 6080 search results of which we consider the first 260 (ranked
by relevance by Google Scholar). We excluded results\textbf{ }on the
unrelated nested sampling technique for soil measurements by removing
publications by one author from the query -``PC Mahalanobis'' and
further manually removed results. (2) Google Scholar citations of
Skilling's original 2004 Nested Sampling paper were searched with
the search query ``nested sampling''\footnote{\href{https://scholar.google.co.nz/scholar?start=180&q=nested+sampling&hl=en&as_sdt=2005&sciodt=0,5&cites=5274326748389338138&scipsc=1}{exact query link}}
in September 2017. This yielded 420 search results which we all considered.
(3) The NASA Abstract Database System was initially searched with
the query \textquotedbl nested sampling\textquotedbl . This gave
1215 results, many of which are simple applications of nested sampling
without methodological contributions, mostly from astrophysics. We
thus limit our search to the arXiv classes \texttt{(arxiv\_class:\textquotedbl stat.{*}\textquotedbl{}
OR arxiv\_class:\textquotedbl math.{*}\textquotedbl{} OR arxiv\_class:\textquotedbl physics.{*}\textquotedbl}),
where papers developing statistical methods and algorithms are (cross-)posted.
This yields 78 results, all of which were considered. (4) Works previously
known to the first author were also included. Out of those four queries
we consider works with novel contributions to any aspect of nested
sampling methods. 

Some restrictions are necessary to focus the content. Firstly, we
limit ourselves to inference problems over continuous parameter spaces.
NS does not require the space to be continuous, only that a prior
is defined from which can be sampled. The ``objects'' considered
can be of categorical nature or of varying dimensionality \citep[e.g.,][]{Brewer2014}.
Furthermore, the review does not go into depth on probability theoretical
analyses of NS. This is mathematically involved and already covered
elsewhere, to which we refer the interested reader in sections \ref{subsec:nsstats}
and \ref{sec:formal}. We exclude works which merely apply nested
sampling in a previously published form to a new problem without modifications
of any aspect of an existing algorithm, and further exclude works
which do not describe the specific method they use.

The review made clear that NS is developed in communities with limited
communication. The first sections introduce NS from the view points
of Bayesian practitioners (\ref{subsec:intro-practical}), statisticians
(\ref{subsec:nsstats}), physicists (\ref{subsec:nsphysics}) and
computer scientists (\ref{subsec:NStree}). Using the language of
each group these sections attempt to allow experts to exchange their
ideas better. The focus of this review are techniques for implementing
the components of NS, enumerated in section \ref{subsec:Components},
in such a way that they efficiently address the PMDIT challenges.
Section~\ref{sec:formal} gives an introduction to the integration
procedure, and references convergence proofs. Termination criteria
are discussed in §\ref{sec:Termination-criteria}, followed by a discussion
of the computational complexity in §\ref{subsec:Scaling}. Diagnostics
to determine the correctness and quality of a NS run are presented
in §\ref{subsec:Correctness-Diagnostics}, including a new test. Sampling
methods for use inside NS are extensively reviewed, including methods
based on random walks (§\ref{subsec:Local-Step-Algorithms}), rejection
sampling (§\ref{subsec:Region-Sampling-Algorithms}), and hybrids
(§\ref{subsec:Hybrid-methods}). Section~\ref{sec:Variations-Integration}
then discusses variations of NS, which soften the hard likelihood
constraint, vary the number of live points and parallelise the algorithm.
In §\ref{sec:Numerical-experiment}, a simple numerical experiment
demonstrates some of the behaviours of NS implementations.

The survey of techniques presented helped inform design decisions
for our own open-source NS implementation. \emph{UltraNest} \footnote{\href{https://johannesbuchner.github.io/UltraNest/}{https://johannesbuchner.github.io/UltraNest/}}
is a high-performance general purpose nested sampling library for
models written in the Python, C, C++, Fortran, Julia or R programming
languages, with a focus on reliability. We document the design decisions
for \emph{UltraNest} in the relevant sections.

\section{Introduction to Nested Sampling}

\label{sec:Introduction}We present introductions aimed at different
audiences, with the goal of enabling different groups to translate
between their languages used. The sections present an introduction
NS from the perspective of a Bayesian practitioner (§\ref{subsec:intro-practical}),
theoretical statistician (§\ref{subsec:nsstats}), physicist (§\ref{subsec:nsphysics})
and computer scientist (§\ref{subsec:NStree}). The components of
NS are then identified in §\ref{subsec:Components}.

\subsection{Conceptual introduction}

To get started, a reference version of the algorithm is introduced.
The theoretical background, justifications and variations are discussed
in subsequent sections. While the algorithm is not limited to continuous
priors, to understand the concepts, it can help to first consider
a parameter space with uniform priors. For example, $V=\int\pi(\theta)\,d\theta$
can intuitively be associated with a volume. Some implementations
also prefer this approach, and support nonuniform, including dependent,
priors by inverse transforming with the cumulative prior distribution
(see Appendix~\ref{sec:Non-factorized-priors}).

\label{subsec:intro-practical}

\begin{figure*}
\includegraphics[width=1\textwidth]{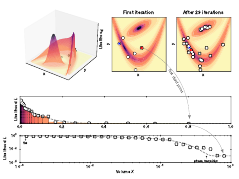}

\caption{\label{fig:progressionillustration}\emph{Top left}: A complicated
likelihood function is defined over a two-dimensional parameter space.
NS begins by evaluating $N=5$ random points. \emph{Top center}: Each
live point (circle) defines a likelihood contour. The lowest likelihood
point (red cross) becomes a dead point. It is replaced by a new, live
point (blue circle), sampled randomly from the prior but above the
contour defined by the dead point. \emph{Top right}: After a few iterations,
the live points concentrate in a small volume at the likelihood peak.
\emph{Bottom} \emph{panels} (top is linear, bottom is logarithmic):
For each iteration, a dead point is placed with its likelihood and
the prior volume estimated by geometric shrinkage. The prior volume
of the sequence shrinks exponentially from right to left. Vertical
bars represent the likelihood shell removed, and are coloured consistent
with the contours shown in the other panels. The bar area is the posterior
weight, and the sum of the bars gives the marginal likelihood $Z$.
The gray dashed curve indicates the true volume-likelihood relation
for this function. In the bottom log-log plot, the phase transition
is marked, which corresponds to the transition from the wide, shallow
yellow regions to the high and steep orange regions in the upper panels.}

\end{figure*}

NS is an integration algorithm that provides both the posterior samples
and the marginal likelihood $Z$. The approach is akin to Lebesgue
integration, which requires keeping track of the height (the likelihood)
and the volume. Lets consider that we want to compute the marginal
likelihood over a $d$-dimensional continuous parameter space. Figure~\ref{fig:progressionillustration}
illustrates the procedure described below.

\paragraph*{Initialisation}

Sample randomly from the prior $N$ live points (e.g., $N=400$) and
evaluate the likelihood function at each point.

\paragraph*{Shrinkage}

Remove the live point with the lowest likelihood, $L_{1}$ (the worst
fit), which becomes the first dead point. Considering that each point
represents $1/N$ of the total volume, this reduces the volume by
a factor of approximately $\delta V=1/N$. Three more estimators of
the removed volume are common: Considering that the samples split
the volume in a uniformly sampled fashion by the sampled $L$ thresholds,
the volume $\delta V$ of the removed shell, $L>L_{\mathrm{min}}$,
is a random variable drawn from a $\mathrm{Beta}(1,N)$ distribution.
This distribution can be randomly sampled, or the geometric mean $\delta V=1-\exp\left(-\frac{1}{N}\right)$
and arithmetic mean $\delta V=\frac{1}{N+1}$ considered as estimators.
This is discussed further in §\ref{subsec:Estimators-and-convergence}.
For all but the smallest $N$, the discrepancy between these estimators
is negligible in practice. For pedagogical simplicity, $\delta V=\frac{1}{N}$
is adopted in this work. With one point removed, the remaining volume
is $V=1-\frac{1}{N}$, i.e., the volume shrank by the factor $(N-1)/N$.

\paragraph*{Likelihood-restricted prior sampling (LRPS)}

A new, independent live point is sampled randomly from the prior,
but it is required that its likelihood exceeds $L_{1}$. This step
is called likelihood-restricted prior sampling, LRPS, also known as
constrained sampling or constrained simulation. Section~\ref{sec:Likelihood-restricted-prior-samp}
extensively discusses LRPS methods. Any region with likelihoods below
$L_{1}$ is not considered any further, and we have again $N$ live
points within a volume. 

\paragraph*{Iterations}

We repeat replacing live points (shrinking and LRPS steps), which
continuously increases the likelihood threshold and shrinks the volume
by approximately a constant factor. Put in another way, removing the
lowest point of $N$ automatically chooses likelihood thresholds such
that the volume decreases by a constant factor, at least on average.
NS scans the prior from the worst to best likelihood. The progression
by constant shrinkage factors reduces the remaining volume exponentially.

\paragraph*{Termination}

After $i$ iterations the remaining volume is exponentially small,
$V_{i}=\left(1-\frac{1}{N}\right)^{i}$, with a high likelihood threshold
selecting live points close to the best-fit parameter peak(s), i.e.,
the likelihood of the remaining points is flat. Further $\Delta V_{i}\times L_{i}$
contributions to $Z$ are thus negligible and the integration can
be stopped. The algorithm has converged in the sense that iterating
further would not significantly alter the result. Other termination
criteria are discussed in §\ref{sec:Termination-criteria}.

\paragraph*{Integration}

Removing a point at iteration $i$ reduced the volume $V_{i}=(1-\frac{1}{N})^{i}$
by $\Delta V_{i}=V_{i}-V_{i-1}=(1-\frac{1}{N})^{i}\times\frac{1}{N}$.
This can be envisioned as a shell of prior volume being peeled off.
The ``level height'' for this integration contribution is just the
likelihood of the dead point, $L_{i}$. Accordingly, each dead point
is assigned the unnormalised weight $\Delta V_{i}\times L_{i}$, and
the integral $Z$ is simply: $Z\approx\sum_{i}\Delta V_{i}\times L_{i}$,
with an error estimate available. The remaining live points at termination
can also be included, with their likelihoods multiplied by the remaining
volume distributed equally among them, $V_{i}/N$. The weighted dead
points are approximate samples from the posterior, and can for convenience
be resampled proportional to $\Delta V_{i}\times L_{i}$ into unweighted
posterior samples.

\paragraph*{Properties}

Figure~\ref{fig:progressionillustration} illustrates some important
aspects of this procedure. Firstly, NS defines an initialisation and
termination point, and can thus run without the user in the loop.
It performs a global exploration of the parameter space. This makes
NS robust to identify and characterise multiple modes. The scanning
from the entire volume to progressively smaller regions is done by
an ever-increasing $L$ threshold. The shrinkage is done by constant
factors, which means the $L$ level is increased dynamically. This
handles heavy and light tailed distributions equally well. Additionally,
it traverses phase transitions, such as from the yellow plateau to
the orange base of the modes in Figure~\ref{fig:progressionillustration}.
This phase transition is highlighted in the bottom panel of Figure~\ref{fig:progressionillustration},
where first the likelihood stays approximately constant as the volume
shrinks, then increases rapidly, making the $\log X-\log L$ curve
non-concave. 

The volume-likelihood plot in the second-to-bottom panel of Figure~\ref{fig:progressionillustration}
shows that much of the space in this simple example has a very low
likelihood, which contributes little to the integral. NS needs to
traverse this space, sometimes for a long time, until the bulk of
the posterior is reached. 

The geometric discovery of new points is done based on the prior.
The likelihood function is queried as an oracle for a binary decision,
namely, whether a suggested new point is inside or outside. In this
geometric sampling of the prior space, NS does not require likelihood
function gradients, making it easy to integrate with complex, numerical
likelihoods from legacy codes.

Finally, NS terminates unsupervised, when the remaining live points
occupy a tiny prior volume, which contributes vanishingly little to
the integral.

To summarise, NS is an attractive algorithm framework for Bayesian
inference because 
\begin{enumerate}
\item it explores the parameter space globally,
\item it handles multi-modal distributions and phase transitions well,
\item it initialises and terminates at a well-defined point without cumbersome
supervision, and
\item it provides both the marginal likelihood and posterior samples.
\end{enumerate}

\subsection{Viewpoint for theoretical statisticians}

\label{subsec:nsstats}

To solve the $d$-dimensional integral of the joint distribution of
$\theta$ and $D$,
\begin{equation}
Z=\int\cdots\int P(D|\theta)\times\pi(\theta)\,d\theta_{1}\cdots d\theta_{d},\label{eq:zintegral}
\end{equation}
NS transforms it into a one-dimensional integral. The survival function
of a likelihood-restricted prior is \citep{chopin2007comments,Chopin2010}:
\begin{align*}
X(L_{\mathrm{min}}) & =pr\{L(D|\theta)>L_{\mathrm{min}}\}\\
 & =\int_{L(D|\theta)>L_{\mathrm{min}}}\pi(\theta)\,d\theta
\end{align*}
Then a ``sorting'' of the prior space via the likelihood function
is achieved by the inverse:
\begin{equation}
Z=\int_{0}^{1}L_{\mathrm{min}}(X)\,dX\label{eq:z1dintegral}
\end{equation}
At first glance, this conceptual transformation has not achieved anything,
because the relevant multi-dimensional spaces cannot be identified
with $L_{\mathrm{min}}$ in all but the most trivial functions.

Instead, NS chooses the $L$ levels such that the corresponding $X$
can be estimated. Note that the inverse of $X$, $L_{\mathrm{min}}(X)$,
is a monotonically increasing function. It is visualised in the bottom
panel of Figure~\ref{fig:progressionillustration}. Suppose $\theta_{1},\ldots,\theta_{N}$
are i.i.d. samples from the prior, and their likelihood is $L_{1},\ldots,L_{N}$.
By definition, the survival function of these likelihood samples is
$X$. Therefore, the probability integral transform of the samples,
$X(L_{1}),\ldots,X(L_{N})$ is i.i.d. standard uniform distributed.
The implication is that points sampled from the prior generate $L$
levels uniform in the prior volume $X$. Lets assume the samples were
indexed so that $L_{1}=\mathrm{min}\{L_{1},\ldots,L_{N}\}$. Then
by the properties of order statistics of a collection of uniform random
variables, the corresponding $X_{1}$ follows $X(L_{1})\sim\mathrm{Beta}(N,1)$
\citep{Skilling2004}. Section~\ref{subsec:Estimators-and-convergence}
below discusses alternatives for this step with different assumptions. 

Setting $L_{\mathrm{min}}=L_{1}$ and repeating the sampling procedure
with the prior restricted to $L(D|\theta)>L_{\mathrm{min}}$ induces
nested sampling. The recursion tracks an ever-shrinking $X$ with
an ever-increasing likelihood threshold $L_{\mathrm{min}}$. Within
the restricted prior space, $X(L_{2}),\ldots,X(L_{N})$ are also uniformly
distributed, specifically from $X(L_{\mathrm{min}})$ to unity. The
sequence of sampled $X_{i}$ thus has the property $X_{i+1}/X_{i}\sim\mathrm{Beta}(N,1)$,
with $X_{1}=1$. This makes estimators such as $Z=\sum_{i}\left(X_{i-1}-X_{i}\right)\times L_{i}$
computable. Section~\ref{sec:formal} discusses convergence proofs
and construction of unbiased integral estimators for $X$, $Z$ and
$\log Z$. 

Above, only a very brief introduction of the ideas involved in NS
was given. We refer to interested reader to \citet{Chopin2010} and
\citet{Schittenhelm2020}, for more formal introductions, to \citet{2014arXiv1412.6368W}
for an analysis of the Monte Carlo point process occurring in the
sequence of finite ordered points used to track shrinkages by sampling
from one likelihood threshold to the next, and \citet{Salomone2018}
for connections to Sequential Monte Carlo.

The popularity of Markov Chain Monte Carlo (MCMC) makes it worthwhile
to draw comparisons between the two iterative Monte Carlo algorithms.
From a starting point, Random Walk Metropolis MCMC constructs a sequence
of points. For choosing the next point, a proposal or transition kernel
needs to be defined, and the Metropolis acceptance rule either chooses
the proposed point or the current point as the next point, proportional
to the posterior probability ratio. If chains are run infinitely long,
the distribution of chain points converges to the posterior distribution.
The performance of MCMC with finite chains crucially depends on the
transition kernel, and many methods have been proposed. Similarly,
the performance of NS crucially depends on the LRPS, and many methods
have been proposed.

The MCMC and NS algorithms can also be qualitatively compared by their
emergent behaviour in typical applications. MCMC typically exhibits
an initial phase where it attempts to identify the posterior bulk.
In this phase, the posterior density is typically rapidly increasing
by orders of magnitude. This initial phase is followed by exploration
of the posterior, where the chain begins to converge, and the number
of effectively independent samples is proportional to the length of
the chain.

We can also identify three emerging phases the NS algorithm exhibits.
Initially, the volume is large and the live points vary by many orders
of magnitude, including many bad fits, so that the dead points receive
weights $\Delta V_{i}L_{i}$ that are ultimately negligible. Because
the live points vary in their likelihood value by many orders of magnitude,
if the algorithm were terminated in these iterations, all of the posterior
weight would be concentrated in the most likely point found so far
$Z\approx L_{i}V_{i}/N$. Where the volume is still substantial and
likelihoods are high, so that $L_{i}\times V_{i}$ is maximal, the
posterior bulk is reached, which we can identify as a second phase.
Here, multiple points receive comparable weights $V_{i}L_{i}$, i.e.,
the posterior becomes resolved into multiple posterior samples. Because
NS needs to track the shrinkage, it cannot rapidly skip ahead to this
phase like MCMC, and therefore (but see §\ref{subsec:Vary-number-of})
the phase of identifying the posterior bulk can take many iterations
(proportional to the information gain \citep{Skilling2004}). Finally,
NS exhibits a phase where the likelihoods are high and very close
to the maximum likelihood, but the volume has become very small. Therefore,
most points receive a small weight and the posterior bulk has been
passed. Here NS differs from MCMC in that continuing the run does
not linearly increase the effective sample size. Section §\ref{subsec:NStree}
and §\ref{subsec:Vary-number-of} discuss methods for bulking the
posterior samples with additional iterations.

\subsection{Viewpoint for physicists}

\label{subsec:nsphysics}

Many Monte Carlo algorithms stem from analogies to physical systems.
To give an example \citep[from][]{Skilling2012,Habeck2015demonic},
consider several gas particles in a box. The position and velocities
of all particles then completely describe the microstate, or configuration
$\theta$ of the system. If the particles are rolling under gravity
within a (perhaps strangely shaped) basin, the total potential energy
of the system $E(\theta)$ can be identified. This is the analogy
to the negative log-likelihood, $E(\theta)=-\log L$. NS initially
generates random particle configurations. The hottest configuration
(highest $E(\theta)$), with energy $\epsilon$ is selected. New configurations
$\theta'$ accessible with a lower energy state $E(\theta')\leq\epsilon$
than the current energy limit are generated, for example by jittering
the particles.

Iteratively replacing the hottest configuration by a cooler configuration
corresponds to a cooling schedule. Monte Carlo cooling schemes are
known from simulated annealing \citep[e.g.,][]{Kirkpatrick1983} and
parallel tempering \citep{Swendsen1986}. NS differs here by choosing
the cooling schedule adaptively, and that it explores a truncated
basin geometrically rather than a smoothed basin proportionally. During
the cooling process, it can occur that the energy changes very little,
while the volume keeps decreasing, followed by an abrupt change of
behaviour where the energy increases rapidly. Such phase transitions
\citep[see e.g.,][for physics background]{Raghavan1975} are problematic
in simulated annealing because it considers the magnitude of energy
changes. In contrast, because NS progresses with constant speed in
volume and considers only the order of the live points, it traverses
phase transitions without issue \citep{Skilling2004}. For plateaus,
see §\ref{subsec:Inside-out}.

NS considers an isolated system with a maximum energy (microcanonical
view) rather than the ensemble average (canonical view). More explicitly,
\citet{Habeck2015demonic} identified several terms from statistical
mechanics in the NS procedure, as follows: The volume of configurations
$X$, with less energy than a threshold $\epsilon$, is
\[
X(\epsilon)=\int_{-\infty}^{\epsilon}g(E)dE,
\]
where the density of states at energy $E$ is \citep[see also][]{2013arXiv1301.6450C}:
\[
g(E)=\int\delta(E-E(\theta))\pi(\theta)d\theta
\]
In probability terms, $g(E)$ describes the distribution of negative
log-likelihood values marginalised over the prior, with $X(\epsilon)$
its cumulative probability distribution. The logarithm of $X$ can
then be identified as the microcanonical entropy $S_{G}(E)=\log X(E)$,
while the logarithm of $g$ is the surface entropy $S_{B}(E)=\log g(E)$.
A microcanonical temperature can then be defined as $T_{G}=X(E)/g(E)$.
The total energy of all configurations, the partition function $Z$,
is defined as $Z=\int e^{-\beta E(\theta)}\pi(\theta)d\theta$ with
$\beta=1$, which is evaluated by NS as $Z=\int e^{-E}g(E)dE$, i.e.,
over the energy levels. For more details, see \citet{Habeck2015demonic}
and \citet{2013arXiv1301.6450C}.

The generation of new configurations can also be considered in analogy
with physical systems. In this case, each configuration is considered
a particle, which inhabits an energy potential. The acceleration of
a particle in an energy basin can be motivated as in the development
of Hamiltonian Monte Carlo \citep[HMC,][]{Neal2011}. HMC constructs
trajectories using the potential energy, which can be considered Keplerian
orbits \citep{Betancourt2017} of random orientation. NS and HMC analogies
differ in two important ways. HMC trajectories conserve the total
energy, partitioned into potential and kinetic energy. This tends
to explore only a narrow range of potential energies, set by the number
of dimensions, and limits HMC's exploration of new configurations
(such as distant basins). In contrast, NS scans potential energies
from hottest to coolest, and generates configurations at all energy
levels. The second difference is that NS searches for new configurations
regardless of their energy, so long as they fulfil the energy threshold.
Thus, the particles receive no acceleration, and the exploration is
purely geometric. We refer the reader to \citet{nielsen2013nested,Martiniani2014,Habeck2015demonic,Baldock2016}
for formulations of NS based on statistical mechanics, and for billard-like
walks, to §\ref{subsec:Sampling-by-direction}.

\subsection{Viewpoint for computer scientists\label{subsec:NStree}}

In computer science, to quickly narrow down a search space, divide-and-conquer
algorithms such as binary search or k-d trees are frequently employed.
Often, algorithms are closely identified with a specific data structure
that fully represents the state at any time. It can be insightful
to investigate the properties of such data structures. In this section,
we identify such a data structure, and phrase NS as an algorithm operating
on it. The representation makes resuming an existing NS run, parallelisation
and dynamically varying the number of live points trivial, and avoids
a special treatment of the final phase of the algorithm.

\begin{figure}

\includegraphics[width=0.4\textwidth]{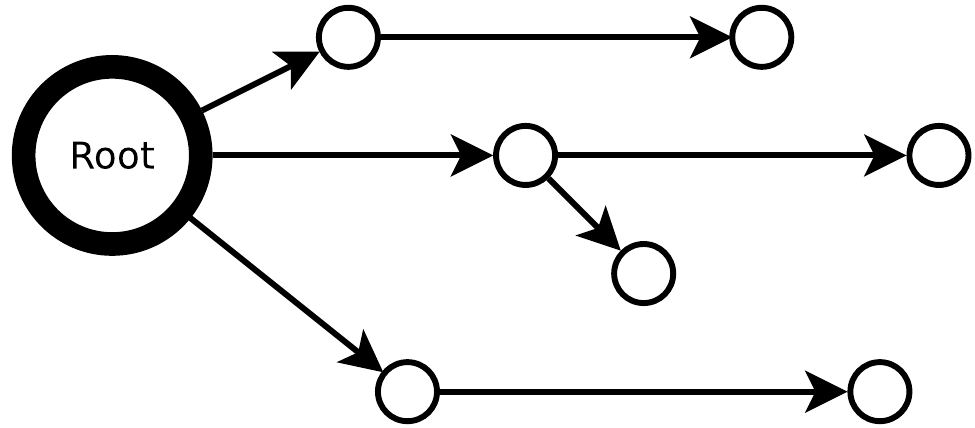}

\caption{\label{fig:NS-tree}NS exploration presented as a tree. The root node
presents the entire volume, with likelihood zero. Each node represents
a point sampled from the prior, restricted to be above the likelihood
threshold of its parent node. In this presentation, the nodes are
also ordered from left to right by likelihood value. Each outgoing
edge splits the volume associated with the parent and donates its
child a volume of $\frac{1}{N}$ where N is the number of parallel
edges.}

\end{figure}

\begin{center}
\begin{figure}

\begin{centering}
\includegraphics[width=0.45\textwidth]{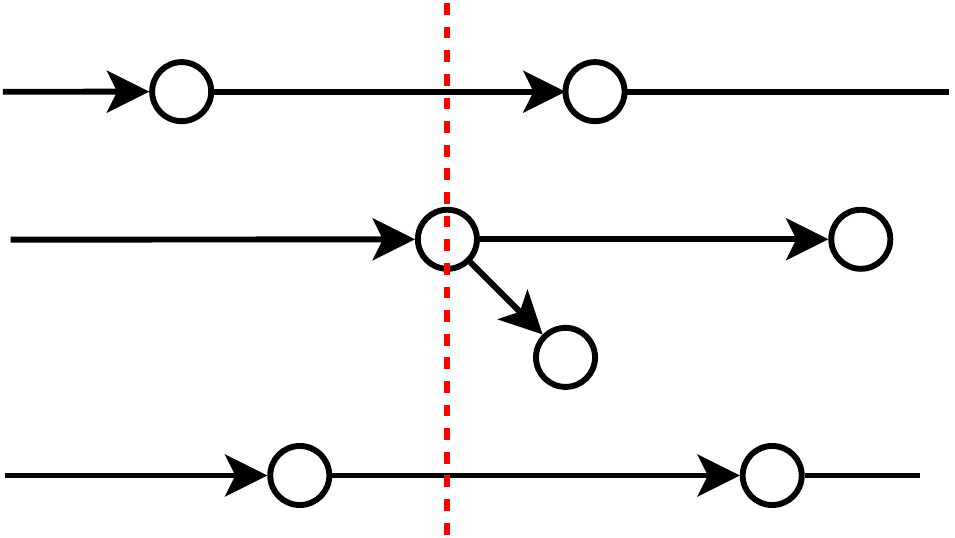}
\par\end{centering}
\caption{\label{fig:NS-tree-search}Search operation: The breadth-first search
iterates in order of likelihood (from left to right). The number of
live points $N$ is the size of the node list, and corresponds to
the number of parallel edges (here: three before the red dashed line,
four after). The volume is reduced, the node assigned a posterior
weight, and the integral updated.}

\end{figure}
\par\end{center}

\begin{algorithm}
\begin{lstlisting}[numbers=left,basicstyle={\scriptsize},tabsize=2]
function NS-BFS:
	given root node (representing the prior volume)
	
	let Q be a list, sorted by likelihood value
	add all children of root node to Q
	
	Z = 0
	V_remaining = 1
	posterior_points = []
	posterior_weights = []
	
	while Q is not empty:
		Nlive = length of Q
		obtain and remove next node from Q
		
		# optional:
		node_expanding_agent(node, Q, Z, V_remaining)
	
		removed_fraction = 1 / Nlive
		remaining_fraction = 1 - removed_fraction
		V_dead = V_remaining * removed_fraction
		weight = V_dead * node.likelihood_value
	
		add node to posterior_points
		add weight to posterior_weights
		Z += weight
		V_remaining *= remaining_fraction

		add all children of node to Q
	return Z, posterior_points, posterior_weights
\end{lstlisting}

\caption{\label{fig:NS-BFS}NS as a breadth-first search algorithm.}
\end{algorithm}

We adapt breadth-first search on a tree structure (see Figure~\ref{fig:NS-tree}).
The search starts from an initial node and keeps a sorted list of
nodes yet to be explored. Initially, the NS tree is a lone root node
representing the full prior volume. Next, we consider sampling $N$
random points inside that volume. This is represented by adding $N$
child nodes of the root node, and illustrated in Figure~\ref{fig:NS-tree}.
We assign the child nodes their likelihood value. A simple breadth-first
search algorithm starts from the root node, keeps a list of ``open''
nodes, and repeatedly explores the one with the lowest likelihood
value until the list is empty. The remaining volume, integrated likelihood
and posteriors are also computed. This is presented in Algorithm~\ref{fig:NS-BFS},
and illustrated in Figure~\ref{fig:NS-tree-search}. 

The formulation in Algorithm~\ref{fig:NS-BFS} is elegant, because
it unifies the nested sampling phase and the remainder integration:
During the remainder integration, none of $N$ parallel nodes have
children, each removed node shrinks that volume by $(N-1)/N$, leading
to equal weights in this phase.
\begin{center}
\begin{figure}

\begin{centering}
\includegraphics[width=0.25\textwidth]{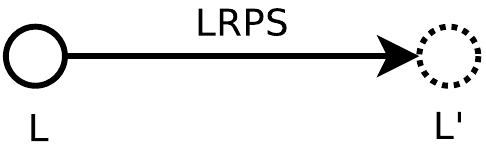}
\par\end{centering}
\caption{\label{fig:NS-tree-insert}Insertion operation: Adding a child node
(right) to the tree means LRPS sampling under the likelihood threshold
of that parent node (left).}

\end{figure}
\par\end{center}

This procedure would soon run out of nodes to explore, so an agent
is needed which adds children to nodes. Figure~\ref{fig:NS-tree-insert}
illustrates that adding a child to a node means LRPS sampling a new
point using the parent's likelihood as the threshold. The simplest
formulation with a constant number of live points adds a single child
node to each node being passed, and stops when the remaining volume
is negligible. This agent is shown in Algorithm~\ref{alg:Simple-exploration-agent}.
In general, an arbitrary number of agents can add children to arbitrary
nodes in the tree.

\begin{algorithm}
\begin{lstlisting}[numbers=left,basicstyle={\scriptsize},tabsize=2]
function classic_node_expanding_agent:
	given node          # current node
	given Q             # current list
	given Z             # current integral estimate
	given V_remaining   # volume left to be explored
	Lmax = (last node in Q).likelihood_value
	frac_remaining = Lmax * V_remaining / Z
	
	if frac_remaining > 0.001 and node has no children:
		new_node = LRPS(node.likelihood_value)
		add new_node as a child of node
\end{lstlisting}

\caption{\label{alg:Simple-exploration-agent}Exploration agent corresponding
to constant-$N$ nested sampling.}
\end{algorithm}

This formulation implies that resuming an interrupted run is trivial,
if the tree is kept. Also, several NS runs as specified in Algorithm~\ref{fig:NS-BFS}
and \ref{alg:Simple-exploration-agent} can run independently in parallel.
For merging, all that is needed is to merge the root nodes, and run
only the integration of Algorithm~\ref{fig:NS-BFS} for the final
result. The formulation also suggests a procedure for converting previously
sampled points with their thresholds into a tree: a LRPS sampled point
can be attached as a child to nodes with likelihood above or equal
the used sampling threshold.\clearpage{}

\subsection{Components of nested sampling implementations}

\label{subsec:Components}\begin{wrapfigure}{R}{0.5\textwidth}%

\begin{centering}
\includegraphics[width=0.4\textwidth]{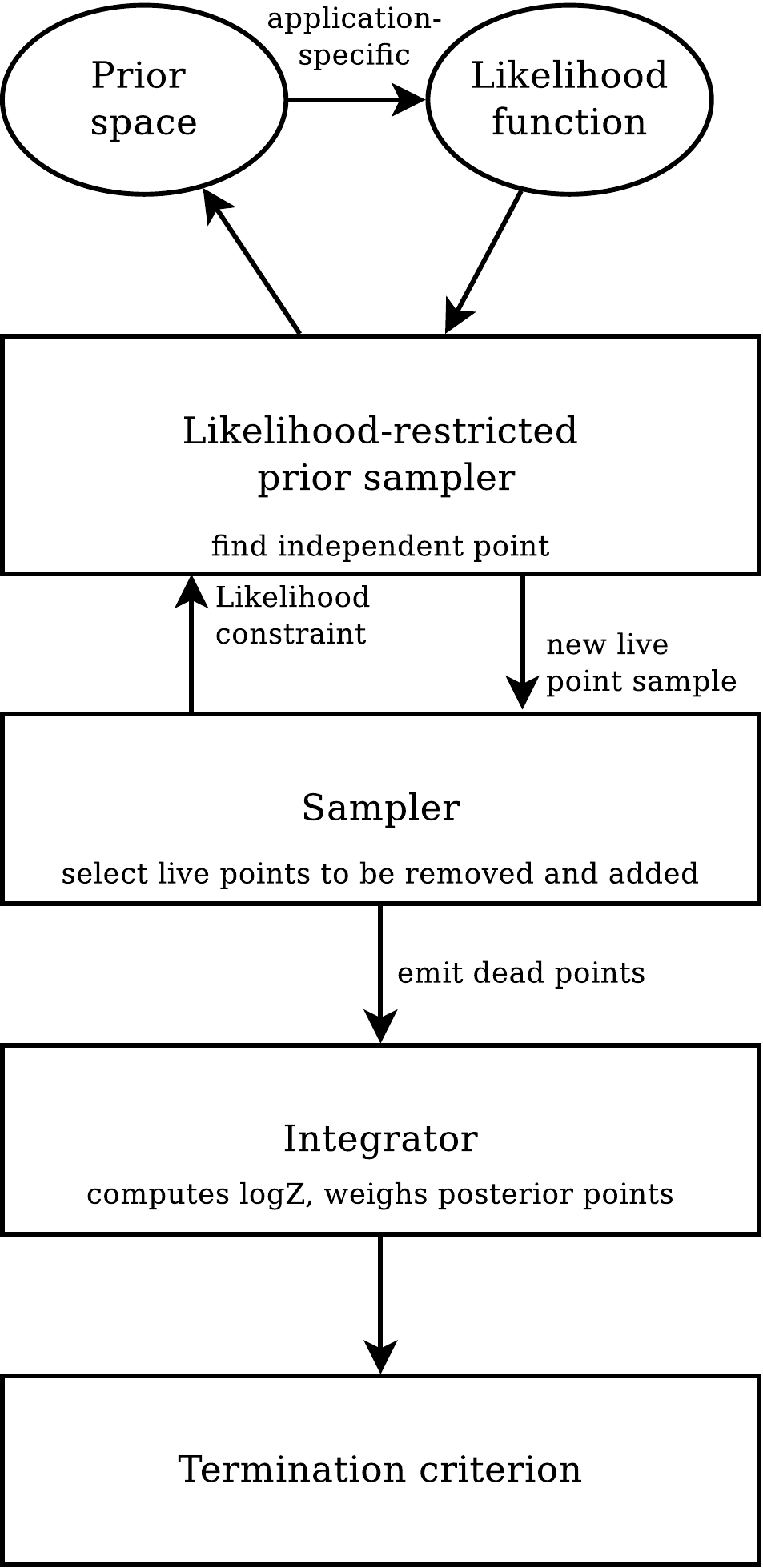}
\par\end{centering}
\caption{\label{fig:nsoverview}Components of a nested sampling implementation.}
\end{wrapfigure}%

To discuss NS in a structured fashion, we identify components of nested
sampling implementations. These are illustrated in Figure~\ref{fig:nsoverview}.
The core (centre) is a NS sampler which keeps a set of live points,
and the likelihood constraint defined by the most recent dead point.
At each iteration, the lowest live point may be replaced by a LRPS,
which uses the application-specific likelihood function and prior
space definition provided by the user. The dead point is passed to
the NS integrator, which weighs these dead points to form a posterior
sample, and estimates the marginal likelihood $Z$ (see formulas in
§\ref{sec:Introduction}). The following sections discuss the individual
components, including the integrator (§\ref{sec:formal}), the termination
criterion (§\ref{sec:Termination-criteria}), the LRPS variants (§\ref{sec:Likelihood-restricted-prior-samp})
and samplers (§\ref{sec:Variations-Integration}).

\section{Integration}

\label{sec:formal}

\subsection{Theory}

NS was introduced in \citet{Skilling2004,skilling2006nested}. Convergence
and unbiasedness was discussed and proven in \citet{evans2007discussion,Chopin2010,skilling2009nested,Keeton2011}.
Making advances to analyse NS theoretically has been the focus of
a few publications \citep[e.g.,][]{khanarian2013nests}. We refer
the interested reader to \citet{2014arXiv1412.6368W}, which makes
connections to the mathematical theory of rare event simulation and
the last particle algorithm, and \citet{Salomone2018}, which draws
connections to Sequential Monte Carlo, nearly considering NS as a
special case of that framework. \citet{Birge2012} presents a generalisation
of NS and connects it to path and bridge sampling, while \citet{Polson2014}
discusses connections to, among others, slice sampling. \citet{chopin2007comments}
and \citet[appendix C]{Feroz2013} prove that posterior samples from
nested sampling approximate the true posterior for continuous and
discontinuous functions, respectively.

\subsection{Estimators}

\label{subsec:Estimators-and-convergence}

\begin{figure}
\begin{centering}
\includegraphics[width=0.5\textwidth]{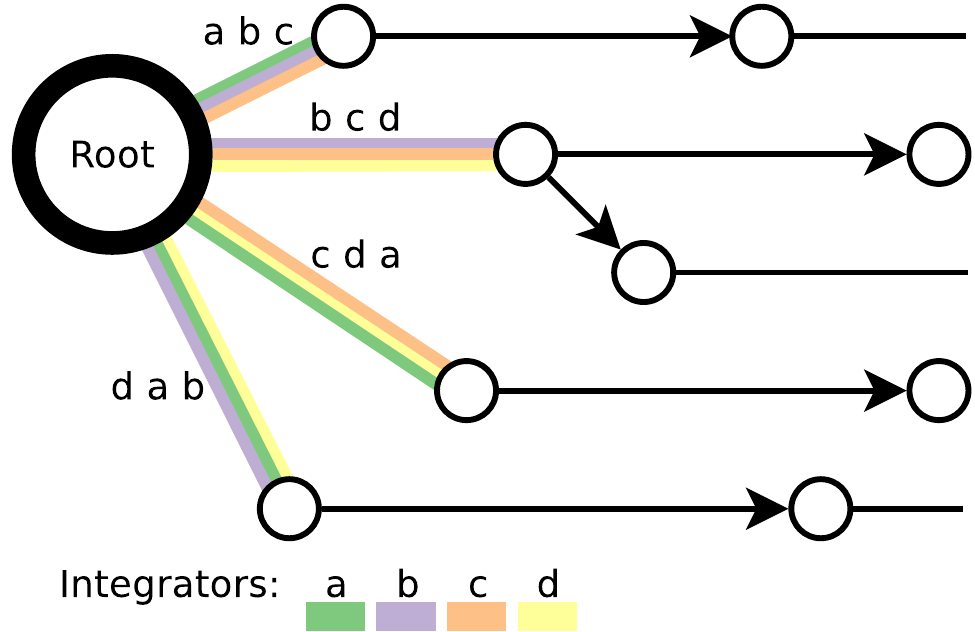}
\par\end{centering}
\caption{\label{fig:NS-tree-bootstrap}NS error estimation with K-fold cross-validation:
Four integrators are invoked, which each are made oblivious to some
of the root children (colours). The spread of integration results
provides an error estimate.}
\end{figure}

NS rests on being able to estimate the volume (shrinkage) at each
iteration, and that the LRPS samples faithfully. LRPS issues are discussed
in §\ref{sec:Likelihood-restricted-prior-samp}, while for the discussion
here, it is assumed that the LRPS is sampling perfectly. This case
is called the idealised algorithm in \citet{Guyader2011}, and perfect
nested sampling in \citet{Higson2017}. 

NS rests on being able to estimate the volume (shrinkage) at each
iteration. Only in special circumstances the volumes are known precisely
\citep{Chopin2008}. In general they can only be estimated. Casting
the volume shrinkage as a Poisson process yields the volume shrinkage
estimator $\delta V=\frac{1}{N}$ \citep{Huber2010,Guyader2011}.
Based on this approach, \citet{Walter2015} derived an unbiased estimator
of $Z$. \citet{chopin2007comments,evans2007discussion} and \citet{skilling2009nested}
discussed whether the ultimate goal is to obtain an unbiased estimator
with minimal variance of $\log Z$ or $Z$ \citep[see also][]{Keeton2011}.
\citet{skilling2009nested} argued that because Bayes factors and
posterior odds ratios interpreted on log-scales are the goal, unbiased
estimators of $\log Z$ should be sought.

Integral estimators rely on estimating the compression ratio at each
iteration. In the order statistics approach of \citet{Skilling2004},
they obtain $V_{i+1}/V_{i}\sim\mathrm{Beta(N,1)}$, and estimate a
geometric log-volume progression as $\log V_{i+1}/V_{i}=-1/N$. The
mean of the Beta distribution gives $\delta V=\frac{1}{N+1}$, which
differs from the unbiased estimator above. Even if not optimal, a
slightly biased estimator is suitable for Bayesian inference if its
bias is negligible relative to its root-mean-square error. This is
the case for the above estimators and their differences, when $N\gg1$,
e.g., $N=100$ \citep{Salomone2018}. For theoretical and online error
analyses of the estimators, we refer the reader to the above works.

The termination of the iterative procedure can also induce a bias
\citep{Walter2015}. \citet{Walter2015} explores the use of families
of estimators, each corresponding to a different random termination,
to remove this bias.

The finite resolution of the $N$ live points leads to a noisy exploration
and likelihood fluctuations \citep{Higson2018}. This can be practically
and elegantly addressed by sub-sampling \citep[first proposed by][]{Higson2018}.
In the tree formulation of §\ref{subsec:NStree}, some of the root
children can be unlinked in a bootstrapping or K-fold fashion. This
is illustrated in Figure~\ref{fig:NS-tree-bootstrap}. In the tree
search formulation one could have several integrators which see only
some of the root children, and use the spread of integration results
(posterior and $\log Z$) to measure the uncertainty. The shrinkage
volume can be randomly generated at each step (using a $\mathrm{Beta}(1,N)$
distribution, \citealp{skilling2006nested}), which then also leads
to a distribution of estimators. Incorporating the two sources of
variance with sampling yields realistic uncertainty also for single
NS runs \citep{Higson2018}.

\subsection{Termination criteria}

\label{sec:Termination-criteria}

The question arises when to terminate nested sampling. In the tree
formulation (§\ref{subsec:NStree}), this means when agents should
stop inserting new children nodes into the tree.

Monte Carlo integration techniques have intrinsic limitations in integrating
black-box functions. The spike-and-slab problem illustrates this.
In Figure~\ref{fig:progressionillustration}, we illustrate a small,
high peak (spike) and a wider ridge (slab). If the spike is very narrow,
it can go unseen by random samples, which will focus on the wider
slab. However, if the spike is very high, it will be crucial for the
integral. As John Skilling puts it: ``It is impossible to find a
flag pole in the Atlantic ocean''. NS integration with sparse sampling
may pass the likelihood threshold that separates the spike and slab
without ever placing live points into the spike. Thus, a spike will
likely go undetected if it is smaller than $V_{i}/N$ \citep{partay2010efficient}.
However, if the spike is located on the peak of the slab, then NS
will find it, unless it terminates too early. 

For less perverse likelihoods, sensible termination can be determined
at runtime. \citet{skilling2006nested} proposed simultaneously estimating
the information gain ($H$, also employed in an error estimate) during
the run and estimating the minimum number of iterations needed to
pass the bulk of the posterior as $HN$. Similarly, the sample entropy
was considered as a termination criterion. Alternatively, \citet{skilling2006nested}
suggests comparing the dead point integral $Z_{i}=\sum_{i}L_{i}\delta V_{i}$
with the largest possible live point contribution $Z_{\mathrm{live}}=L_{\mathrm{max}}\times V_{i}$,
and terminating when the ratio becomes very small, $Z_{\mathrm{live}}/Z_{i}<\epsilon$,
with $\epsilon\ll1$, for example $\epsilon=10^{-3}$. This is also
the agent behaviour implemented in Algorithm~\ref{alg:Simple-exploration-agent}.
This can be further refined with $Z_{\mathrm{live}}=\frac{V_{i}}{N}\sum_{j=1}^{N}L_{j}$,
trapezoid rule integration, or bootstrapping, but in practice is not
more reliable. 

A variety of other termination criteria have been considered. For
example, \citet{schoniger2014model} proposes terminating when the
LRPS becomes extremely inefficient. Low efficiencies caused by complex
degeneracies, can indicate that the model could benefit from reparametrisation
\citep[see][for a similar situation with MCMC]{Papaspiliopoulos2007}.
\citet{Baldock2016} suggests monitoring the temperature (HMC momenta,
see below §\ref{subsec:Sampling-by-direction}). 

The problem of yet unidentified, hidden peaks cannot be addressed
in a general and reliable way with information available during the
run. Therefore running a few iterations longer than seemingly needed
is most effective in practice. This is what the $\epsilon$ remainder
fraction criterion effectively does.

Termination can also be addressed with domain knowledge and reparametrisation.
Some likelihoods have absolute upper bounds. For example, in a Gaussian
likelihood with measurements $d_{i}$ with fixed uncertainties $\sigma_{i}$
fitted with an arbitrary model $m$, the weighted sum of squared deviations,
$\chi^{2}$, is positive: $-2\times\log L=\chi^{2}=\sum_{i}\left(\frac{m(i,\theta)-d_{i}}{\sigma_{i}}\right)^{2}>0$.
The likelihood bound directly gives upper bounds on $Z_{\mathrm{live}}$
of future iterations.

\subsection{Complexity scaling}

\label{subsec:Scaling}

We can now consider the computational complexity of NS integration.
This depends on (1) the information gain of the posterior compared
with the prior (I), which determines the shrinkage necessary to reach
the bulk of the posterior, (2) the number of live points, which determines
the shrinking per iteration, and (3) the computational complexity
of the LRPS per NS iteration to find a reliable new point, which is
subject to peculiar degeneracies (P), multi-modality (M) and dimensionality
(D) issues.

The NS complexity scales linearly with the number of live points $O(N)$,
due to the slower shrinkage \citep{Skilling2004}. Beyond this, the
scaling of the LRPS method can be arbitrarily hampered by complex
posterior shapes that need to be navigated until a new independent
sample is obtained. Assuming the latter scales linearly with dimensionality
$d$, \citet{skilling2009nested} gives the cost scaling of NS as
$O(d^{2})$. In practice, \citet{Handley2015a} demonstrate a scaling
of $O(d^{3})$ for their implementation. This makes NS perhaps less
attractive for very high-dimensional problems with many thousands
of parameters such as fitting large hierarchical Bayesian models or
neural networks. However, \citet{Javid2020a} demonstrate that neural
networks can be fitted, and NS is attractive because the global exploration
avoids choosing overly complex networks.

However, the relation between cost and live points is more complex.
While convergence slows with the number of live points, some LRPS
methods work more efficiently the more live points, as they help map
out the likelihood constraint and can identify the approximate neighbourhood
where new points are likely successful (see §\ref{sec:Likelihood-restricted-prior-samp}).
Following \citet{Allison2014}, this is illustrated with the ellipsoidal
rejection sampling technique (discussed further in §\ref{subsec:Region-Sampling-Algorithms})
for the case of ellipsoidal, mono-modal likelihoods. For this, the
LRPS cost per iteration is empirically found In Appendix~\ref{sec:Ellipsoidal-sampling-efficiency}
to scales as:

\begin{equation}
O(C_{\mathrm{ell}})=\exp\left\{ \left(\frac{6.83\times d^{1.9}}{N}\right)^{3/4}\right\} \label{eq:empiricalscaling}
\end{equation}
The exponential increase becomes crucially important when the number
of live points is small. This likely encodes the curse of dimensionality,
and the inherent limitations of rejection techniques. Equation~\ref{eq:empiricalscaling}
suggests that the number of live points for ellipsoidal sampling should
not be lower than $7\times d^{2}$. 
\begin{figure*}
\includegraphics[width=1\textwidth]{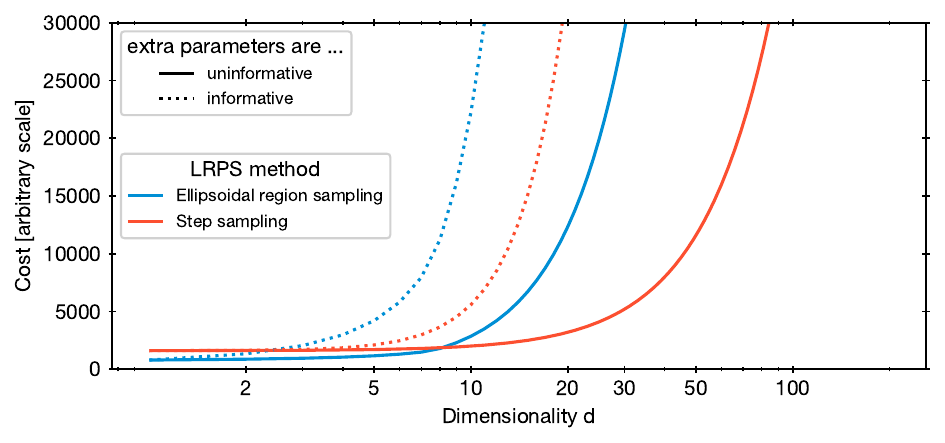}

\caption{\label{fig:ellipsoidalcostscale}Number of likelihood evaluations
on Gaussian-like likelihood functions. The blue curves show two cases
with different information gain changes with dimensionality. Both
cases indicate a low cost at $d<30$ dimensions for ellipsoidal NS
under optimal conditions (likelihood with ellipsoidal contours). Red
curves illustrate the shallower, polynomial scaling of step samplers.}
\end{figure*}

The total cost is obtained from the per-iteration cost and the number
of iterations needed. Shrinking from the prior volume $V_{p}$ until
a low fraction $\epsilon$ of a target posterior volume $V_{t}$ requires
$i=N\times\ln\frac{V_{p}}{V_{t}\times\epsilon}$ iterations (see §\ref{subsec:intro-practical}).
The ratio $H=\ln\frac{V_{p}}{V_{t}}$ here is the information gain
from the prior to the posterior. Combining the acceptance rate formula
of eq.~\ref{eq:empiricalscaling} and the number of iterations $i$,
and the sampling of the initial $N$ live points, \citealp{Allison2014}
obtain the total nested sampling cost as:
\begin{equation}
C=N+i\times C_{\mathrm{ell}}=N+N\times\ln\frac{V_{p}}{V_{t}\times\epsilon}\times C_{\mathrm{ell}}\label{eq:nscost}
\end{equation}
The factor $V_{p}/V_{t}$ is problem-specific. It is not easy to study
the scaling with dimensionality, as varying the dimensionality implies
analysing a different problem. Here we consider two cases: (A) If
new parameters are added, and they are all updated with the same information
gain, then $V_{p}/V_{t}=K^{d}$. This increases the cost of $C$ to
$O(N\times C_{\mathrm{ell}}\times d)$. (B) If adding new parameters
only redistributes the same information, then $V_{p}/V_{t}=K$ remains
constant with dimensionality, and thus the cost is of order $O(N\times C_{\mathrm{ell}})$.
For these two cases, the total costs $C$ are plotted in Figure~\ref{fig:ellipsoidalcostscale},
with $N=400$. The normalisations are arbitrary and cannot be compared.
Ellipsoidal NS clearly works best in $d<30$ dimensions. This agrees
with our practical experience with the MultiNest ellipsoidal sampling
implementation, which begins to break down close to that dimensionality.
For comparison, the aforementioned $d^{2}$ scaling of a MCMC LRPS
algorithm is also shown as red lines, with an arbitrarily chosen base
cost added. Interestingly, the above analysis suggests that for ellipsoidal
problems, a $d^{2}$ scaling is possible with ellipsoidal sampling
if the live points are increased quadratically. However, the optimistic
case where the likelihood indeed has elliptical contours was analysed,
while for other cases, the cost may be higher.

In practice, the computational cost is specific to the model structure
as well. Thus numerical testing is required (see for example Figure
15 in \citet{Pitkin2017} and Figure 6 in \citealp{Trassinelli2020}).
Finally, the discovery of likelihood peaks is also regulated by $N$
(see §\ref{subsec:Scaling}). Because of this, running NS with very
small $N$ does not necessarily give useful ``quick look'' results.

\subsection{Correctness diagnostics}

\label{subsec:Correctness-Diagnostics}

How can a complete NS implementation (including integrator, LRPS,
sampler, termination criterion) be evaluated for correctness? We can
divide in two categories of tests, following \citet{stokes2016equidistribution}.

\subsubsection{Outside-in: Application to known functions}

Firstly, the NS implementation can be applied to test problems with
known properties. The simplest case is likelihood functions where
the integral $Z$ is analytically known \citep[e.g.,][]{Preuss2007}.
The main limitation here is that analytic likelihood functions often
do not resemble real-world problems. A further problem is that often,
for example, when choosing a Gaussian likelihood expression, the integral
is dominated by a narrow range of likelihood values. Thus, the LRPS
and integration are therefore potentially only tested on a low number
of iterations. This can be improved by choosing heavy-tailed distributions.
An extreme case is $L=\min(\theta^{-1},e^{100})$ with $\theta$ defined
over the unit interval, which makes all dead points yield approximately
equal weights.

Secondly, the LRPS can be tested in isolation. \citet{Buchner2014stats}
proposed a LRPS test that verifies that the shrinkage caused by the
LRPS behaves as expected ($1-1/N$). This is based on likelihood functions
where one can analytically compute the volume enclosed at a given
likelihood $X(L)$. For example, in a Gaussian likelihood, the circular
likelihood contours can be identified with an ellipsoid. Then, from
a sequence of likelihoods obtained from repeated LRPS, volume shrinkages
can be computed and compared to expectations. If the likelihood of
the found point is systematically lower or higher, then the LRPS is
noticeably incorrect. This can be applied to many posterior shapes,
including multi-modal Gaussians. A particularly sensitive test is
the hyper-rectangle $L=\max_{i}||\theta_{i}-\frac{1}{2}||^{-1}$,
because its shape is far from Gaussian and exhibits many corners in
high dimensions. Importantly, this test is independent of the tail
weight, as the likelihood only enters in NS weighting of points. It
can also be applied in very high dimensions to tune LRPS parameters.

\subsubsection{Inside-out: Diagnostics at or after run-time}

\label{subsec:Inside-out}

The second group of tests tries to notice during the NS run when assumptions
are broken.

Likelihood functions with plateaus can cause problems in nested sampling
\citep{skilling2006nested,Schittenhelm2020}. This is because the
ordering of the prior space is not available, and a large volume is
associated with a vanishing likelihood interval in eq.~\ref{eq:z1dintegral}.
Therefore, if two live points have the exact same likelihood, this
should cause alarm. To address this, \citet{Fowlie2021} proposed
a small modification to the nested sampling algorithm iterations to
remove all live points with $L_{\mathrm{min}}$ without replacement
before replenishing the live point set to size $N$.

One assumption is that the LRPS samples correctly according to the
constrained prior. \citet{stokes2016equidistribution} proposed tests
for uniformity in 2-dimensional problems. Firstly, they count the
empty cells in a segmentation, and compare them with a Bayes factor
to uniform expectation. Secondly, they develop an equi-distribution
test that measures the concentration of samples with an entropy, and
compares that to a uniform expectation. Finally, they visualise deviations
from uniformity with quantile-quantile plots. These tests however
appear limited to very low-dimensional problems.

\citet{Higson2018,Higson2019} contributed visualisations of the uncertainty
in the inference, in particular of the posterior distributions. This
is achieved by sub-sampling (see §\ref{subsec:Estimators-and-convergence})
completed nested sampling runs, and plotting the spread of posteriors.
Relatedly, they propose diagrams of volume ($\log X$) vs. parameter
value ($\theta_{i}$), which allows insight into the structure of
the parameter space, and how the LRPS replaces samples.

\citet{Higson2019} further proposes testing the variance between
multiple independent NS runs to the variance expected from sub-sampling
a single run. Here, one can check the expectation for each parameter
$\theta_{i}$, a combination $f(\theta)$, or $\log Z$. Excess variance
can occur when the LRPS is sampling imperfectly, and its samples and
shrinkages are correlated. \citet{Higson2019} also considered the
expectations between only two runs, when the computation is very costly.

Finally, \citet{Fowlie2020} pointed out that if the samples returned
from the LRPS are independent and perfectly distributed according
to the constrained prior, then where they are inserted into the sorted
live points list should be uniformly distributed. They proposed an
insertion order test to check this condition. The insertion order
of each new sample is collected. The order distribution is tested
with a Kolmogorov-Smirnov (KS) test every $N$ iterations, and for
the full run. \citet{Fowlie2020} demonstrates that this works well
to detect problematic runs in practice on toy problems. Alternatively,
the insertion order test could also be used as a quality indicator.
When the test triggers, the sample collection is reset. Then the number
of iterations until the test triggers can be interpreted similar to
an auto-correlation length in random walk MCMC algorithms. However,
the samples collection likely still needs to be truncated occasionally,
so that a recent addition of poor samples is not diluted by a preceding
high number of good samples. A limitation here is that the KS test
as typically implemented is only valid for continuous variables. \citet{Fowlie2020}
show that the rounding to integers makes the distribution non-uniform,
and the test is less sensitive than it could be.

We propose an improvement of the power of the test with a statistic
suitable for testing whether discrete numbers are uniformly distributed.
We begin with the Wilcoxon-Mann-Whitney U test \citet{Mann1947},
which tests two sequences of observations, of length $n_{1}$ and
$n_{2}$. For each observation in the first sequence, the number of
smaller and equal observations in the second sequence is recorded
(``wins'', $W$), as well as the number of equal observations (``ties'',
$T$). Then the $U$ statistic is:
\[
U=W+\frac{1}{2}T
\]
Then $U$ is normal distributed with mean
\[
m_{U}=\frac{n_{1}\times n_{2}}{2}
\]
and standard deviation
\[
{\displaystyle \sigma=\sqrt{\frac{n_{1}n_{2}}{12}\left(n+1-\sum_{i=1}^{k}\frac{t_{i}^{3}-t_{i}}{n(n-1)}\right)}}
\]
with $t_{i}$ the number of observations that share order $i$ and
$n=n_{1}+n_{2}$. In other words, $z=\frac{U-m_{U}}{\sigma_{U}}$
is approximately standard normal and should rarely exhibit, for example,
$|z|>3$, a $3\sigma$ excursion.

\begin{table}
\begin{centering}
\hfill{}%
\begin{tabular}{cc|cc}
$N$ & Coverage & KS test & U test\tabularnewline
\hline 
\hline 
1000 & 0.9 & 1.00000 & 0.99723\tabularnewline
1000 & 0.96 & 0.13472  & 0.19984\tabularnewline
1000 & 0.98 & 0.01416 & 0.02627\tabularnewline
400 & 0.9 & 1.00000 & 0.69506\tabularnewline
400 & 0.96 & 0.02344 & 0.04745\tabularnewline
400 & 0.98 & 0.00589 & 0.00926\tabularnewline
100 & 0.9 & 0.07364  & 0.07745\tabularnewline
100 & 0.96 & 0.00988 & 0.00771\tabularnewline
100 & 0.98 & 0.00593 & 0.00383\tabularnewline
\end{tabular}\hfill{}%
\begin{tabular}{cc|cc}
$N$ & Slant & KS test & U test\tabularnewline
\hline 
\hline 
1000 & 0.9 & 0.31010 & 0.44932\tabularnewline
1000 & 0.96 & 0.01864 & 0.02873\tabularnewline
1000 & 0.98 & 0.00505 & 0.00728\tabularnewline
400 & 0.9 & 0.06385 & 0.11478\tabularnewline
400 & 0.96 & 0.00635 & 0.01084\tabularnewline
400 & 0.98 & 0.00343 & 0.00425\tabularnewline
100 & 0.9 & 0.01205 & 0.01653\tabularnewline
100 & 0.96 & 0.00419 & 0.00363\tabularnewline
100 & 0.98 & 0.00335 & 0.00281\tabularnewline
\end{tabular}\hfill{}
\par\end{centering}
\caption{\label{tab:teststats}Performance comparison of KS test and U test.
In the left table, the integers are distributed from $0$ to $N\times\mathrm{Coverage}$,
in the right table they are distributed with probability $p_{i}\propto i^{\mathrm{Slant}}$.
The fraction of tests reporting a $3\sigma$ significance ($p<0.0027$)
is reported on the right. The U test fractions are higher in almost
all cases.}
\end{table}

Lets now imagine that the second sequence is a very large sample of
reference points ($n_{2}=N\times M$ with $M\gg1$). They are uniformly
spread from $0$ to $N-1$, thus each order is represented by $M$
samples. The first sequence of observations are the collected insertion
orders, $O_{i}$. These are integers ranging from $0$ to $N-1$.
In that case, a observed insertion order $O$ will have $M$ ties
and $O\times M$ wins. The test statistic becomes:
\[
U=M\times\sum_{i=1}^{n_{1}}\left(O_{i}+\frac{1}{2}\right)
\]
The mean becomes $m_{U}=\frac{1}{2}\times n_{1}\times N\times M$
while the standard deviation simplifies because $M\gg1$ and $n_{2}\gg n_{1}$,
to ${\displaystyle \sigma=\sqrt{\frac{n_{1}}{12}}\times N\times M}$.
Cancelling out $M$ which occurs in $U$, $m_{U}$ and $\sigma$,
we find that
\[
z=\frac{\left(\sum_{i=1}^{n_{1}}\frac{2O_{i}+1}{N}\right)-n_{1}}{\sqrt{n_{1}/3}}
\]
is standard normal distributed. This is easy to numerically confirm
by sampling $n_{1}$ integers sampled uniformly in $[0,N)$ and plotting
the $z$ values. We make two important remarks: Firstly, in this form,
the $U$ test allows $N$ to vary from iteration to iteration. Secondly,
the sign of the $z$ statistic indicates the direction of the bias.

Now the sensitivity of the KS test and the $U$ test (both two-sided)
can be compared. For reasonable numbers of live points, two scenarios
are considered in Table~\ref{tab:teststats}: In the first simulation
(top), generated insertion orders are truncated to only cover the
range $0$ to $\left\lceil N\times C\right\rceil $, with $C<1$.
In the second simulation (bottom), generated insertion orders are
simulated from a mildly slanted power law distribution, with order
$i$ more likely at the low end. We simulated $100,000$ samples of
size $N=1000,\,400,\,100$ and apply the KS and $U$ tests. The fraction
of tests reporting a $3\sigma$ deviation are compared in Table~\ref{tab:teststats}.
In general, the U test has a higher detection rate. However, when
the detection rate is already high ($>50\%$), or very close to the
expected false positive rate, the KS test sometimes performs similarly
or better. However, these are less interesting edge cases. Additional
to being more sensitive, the U test is slightly simpler to implement,
as only $\frac{2O+1}{N}$ and the number of samples $n_{1}$ need
to be accumulated, instead of entire histograms.

The test can be applied in three different ways: (1) on the full run
as in \citet{Fowlie2020}, (2) every $N$ iterations \citep{Fowlie2020},
probably with a Bonferroni correction, and (3) accumulate until $|z|$
exceeds a predefined threshold. For example, when simulating $10^{7}$
iterations with uniform insertion order, resetting the accumulation
statistic when $|z|>4$ leads to segment lengths no shorter than $10^{5.5}$.
If shorter segments are regularly produced, or more specifically,
if the number of segments of a NS run is larger than the number of
iterations divided by $10^{5.5}$, this is an indication that the
run is biased. This option avoids selecting a chunk size to apply
the test.

\section{Likelihood-restricted prior sampling (LRPS)}

\label{sec:Likelihood-restricted-prior-samp}

LRPS is the crux of NS. Empirical statements about what NS can or
cannot do are at the mercy of the LRPS implementation. This section
tries to convey that a large diversity of solutions are possible and
have been considered.

\begin{figure*}
\begin{centering}
\includegraphics[width=0.8\textwidth]{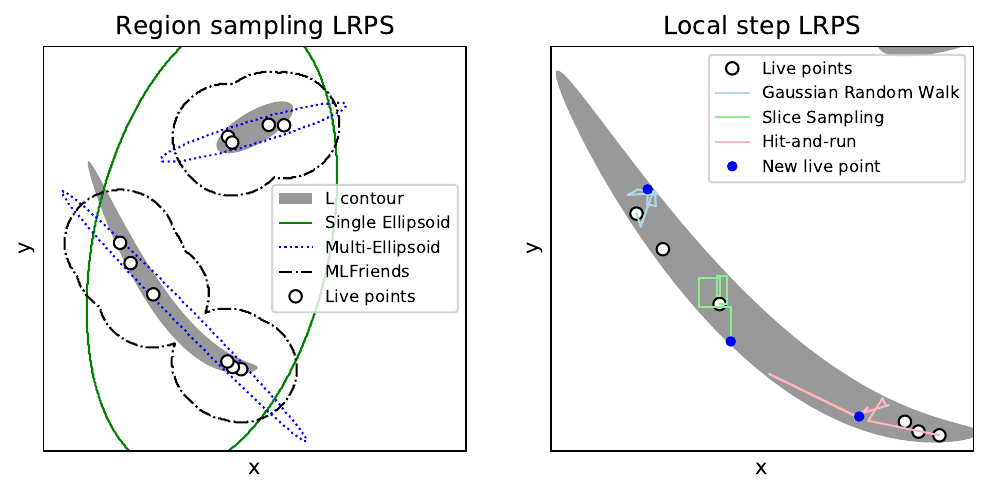}
\par\end{centering}
\caption{\label{fig:LRPSillustration}LRPS methods. At an iteration for the
example from Figure~\ref{fig:progressionillustration}, different
methods for finding a new live point are illustrated. \emph{Left panel}:
Region sampling methods construct regions based on the live points
(white circles) to approximate the unknown likelihood contour (grey).
Rejection sampling based on these contours becomes inefficient if
the contours are too large, and problematic for NS integration if
it misses a region (e.g., the top left tail). The enlargement is intentionally
chosen too small here. \emph{Right panel}: Local step methods start
Metropolis random walks (coloured lines) from a live point (white
circle). Steps outside the likelihood contour are rejected. If the
step proposal is good and the number of steps large, a new point (blue
circle) is reached that is independent of the starting point.}

\end{figure*}

LRPS is supposed to deliver an independent, new point sampled from
the likelihood-restricted prior. This can be difficult to achieve
perfectly. However, in practice, and as elaborated in \citet{Salomone2018}
with theoretical arguments, good NS results can be achieved also with
mildly correlated points.

\subsection{Local step algorithms (MCMC-based)}

\label{subsec:Local-Step-Algorithms}

A new point can be sampled from the prior with MCMC \citep{Skilling2004}.
Typically, either the recently deceased point, or a random point are
chosen as the start of a random walk, or more generally, invariant
move steps with respect to the constrained prior. A new point is proposed,
and accepted if it exceeds the current likelihood threshold. In flat
priors, this makes the random walk a purely geometric exploration.
Such random walks are illustrated in the right panel of Figure~\ref{fig:LRPSillustration}.
The random walk proceeds for a number of steps, after which the final
point is returned to the NS sampler as a (nearly) independent sample. 

Any random walk MCMC solution to LRPS needs (1) a step proposal, (2)
a recipe for adapting the proposal to the continuous shrinking as
NS progresses and (3) the number of steps. Unfortunately, the literature
often lacks these implementation details which severely limits the
usefulness of numerical comparisons. 

\subsubsection{Leveraging live point knowledge}

\label{subsec:Exploiting-live-point}

To craft a good step proposal, two properties can be leveraged: Firstly,
the current live points are already distributed at least approximately
according to the current contour, and trace out the relevant space
and its geometry, if they have been sampled faithfully from the prior
up to the current iteration. This assumption is certainly justified
for the initial live points, which are drawn directly from the prior.
Secondly, the behaviour of the contour changes little from iteration
to iteration, as the volume shrinks by a small fraction, so that very
similar problems have to be solved in sequence. 

The second property is leveraged by reusing optimised proposals between
iterations, for example if the previous iteration's acceptance rate
was low, the proposal is shrunk before use in the next iteration \citep[e.g.,][]{sivia2006data}.
However, such adaptions do not maintain detailed balance. To address
this, \citet{Salomone2018} suggest a warm-up NS run with adaptive
proposals turned on and stored for each iteration, and followed by
a final run that uses these proposals, now without adaptations.

The first property is leveraged by many authors by estimating the
sample covariance from the live points \citep{Veitch2010,schuet2011model}
or determining the principal directions \citep{2009arXiv0912.2317N}
to understand size and orientation of the current space. 

More sophisticated procedures are necessary to handle multi-modality.
\citet{Martiniani2014} pre-computes a database of minima and performs
an exchange move, whereby the MCMC can swap between them. Taking advantage
of the live points, \citet{Handley2015} performs iterative Jarvis-Patrick
clustering \citep{jarvis1973clustering} and then estimates covariances
based on the member points of each cluster, obtaining a local covariance
used for walks started from members of that cluster. MCMC schemes
originally intended for unimodal distributions can be extended to
handle multiple modes by applying a clustering algorithm to the live
points, overlaying the cluster points by subtracting cluster means
from member points, and to derive the proposal from the shifted points
(e.g., their sample covariance). Then the MCMC algorithm begins its
chain from a random live point as its starting point.

\subsubsection{Sampling by vicinity}

The first fully specified proposal was presented in \citet{sivia2006data}.
A multi-dimensional Gaussian is used, starting from the dead point.
If the number of accepts $a$ dominates the number of rejects $r$,
the standard deviation is increased by a factor of $\exp(1/a)$, otherwise
decreased by a factor of $\exp(1/r)$. \citet{brewer2011diffusive}
prefers a heavy-tailed, highly multi-scale proposal to avoid adaptation
at the current iteration. Numerous other MCMC proposal distributions
have been applied \citep[e.g.,][]{liu2016evaluating,beaton2017room,2013ApJ...778...32P}.
A Gaussian random walk is illustrated with blue lines in the right
panel of Figure~\ref{fig:LRPSillustration}.

Another solution comes from \citet{Veitch2010}, who uses a Student-t
distribution with 2 degrees of freedom proposal scaled to 10\% of
the covariance of live points, estimated every 10 iterations. However,
10\% of steps use a proposal inspired by differential evolution: two
additional live points are selected, and the difference vector added
to the current point. These attempt to address multi-modal distributions;
however, they observe that for a specific problem even with 1000 MCMC
chain steps, LRPS artefacts remain. For higher efficiency in intermediate
or high dimensional problems, Gibbs \citep{murray2006nested} and
component-wise proposals along the principal components \citet{2009arXiv0912.2317N}
have been proposed. \citet{2016arXiv161110189T} primarily uses per-parameter
MCMC proposals with a uniform distribution, but also includes a crossover
step to combine live points. This work does not verify the correctness
of their implementation. A later publication by the same author includes
MCMC proposals that do not preserve detailed balance (\citealp{Trassinelli2019},
their first high-failure recovery procedure).

To incorporate knowledge of the problem parameter space geometry,
\citet{Javid2019} introduced proposals along spherical arcs and wrapping
of circular parameters.

\citet{Moss2020} deforms the proposal space to flatten peculiar shapes
and bring multi-modal distributions together. A neural network is
optimised to transform the live points with least information loss
to a standard Gaussian distribution. Sampling from that simpler surface
can be more efficient, and if proposals consider the space compression
by the non-linear flow via the Jacobian, samples from the (restricted)
prior can be obtained. This method appears to be very promising and
general, but also complex to implement and train correctly in practice.

\subsubsection{Sampling by direction}

\label{subsec:Sampling-by-direction}

\citet{Betancourt2011} derived Constrained Hamiltonian Monte Carlo
(CHMC), the straightforward application of Hamiltonian Monte Carlo
to likelihood constrained prior sampling. For simplicity, we describe
its behaviour in a suitable parametrisation where the prior is flat.
Then CHMC makes straight steps of length $\xi$ in a chosen direction
until the step violates the constraint. There, it reflects off the
boundary, i.e. the momentum vector reverses normal to likelihood (constraint)
gradient, and continues. This technique thus requires a step size
and a computable likelihood gradient. Galilean Monte Carlo \citep[GMC;][]{Skilling2012}
adds a reverse step to CHMC upon rejection of the reflection, so that
the chain succeeds more often. This technique is related to reflections
discussed in \citet{neal2003}. Demonic Nested Sampling \citep{Habeck2015demonic}
softens the likelihood contours by storing excess likelihood (energy)
in a demon. This means that a MCMC procedure then tends to turn away
from the border without requiring the use of gradients. When gradients
are available, Demonic Nested Sampling can store the CHMC velocity
vector, and solve a combined momentum-position space. \citet{Baldock2017}
also presents a HMC-based nested sampling extension, and compares
GMC with a CHMC version that stores the momentum of each live point,
finding them to outperform simpler random walk MCMC algorithms \citep[see also][]{nielsen2013nested}.

Another procedure is the sampling from an existing point along a line.
Here, three families of such algorithms are presented, all of which
are available in \emph{UltraNest}:

1) In slice sampling \citep{neal2003}, parameter space axes are iterated
through for the proposal direction. Uniform sampling is achieved by
choosing distant bounds, sampling uniformly between them and shrinking
the bound towards that side for every reject until a sample is accepted
\citep{Handley2015}. The slice sampling random walk is illustrated
with green lines in the right panel of Figure~\ref{fig:LRPSillustration}.
For non-uniform priors, an additional step is needed, by sampling
the height of the prior distribution with an auxiliary variable. For
flat priors, or placing slices through a reparameterised space which
make the prior flat, the likelihood function is used as an oracle
(above the likelihood threshold, or not). The NS literature often
remains vague which exact variant of slice sampling is implemented.
A positive example is \emph{PolyChord}, which proposes along principle
axis after the next \citep{Handley2015}. The principle axis are obtained
from the sample covariance matrix of live point cluster where the
walk has started (see §\ref{subsec:Exploiting-live-point}).

2) In hit-and-run Monte Carlo \citep[HARM,][]{doi:10.1137/1116083,smith1984efficient},
a random direction is instead chosen in each step. The algorithm variant
with bound shrinkage was presented explicitly by \citet{Kiatsupaibul2011}
and is most closely related to slice sampling. Such a walk is illustrated
with pink lines in the right panel of Figure~\ref{fig:LRPSillustration}.
Even in complicated geometries, HARM is highly effective in mixing
and scales well with dimensionality \citep[see][and references therein]{2013arXiv1312.7061C,Kiatsupaibul2011},
\citet{2013arXiv1312.7061C} also show that it outperforms slice sampling.
\citet{stokes2017new} tested this technique for NS with non-convex
surfaces, albeit in low dimensions.

3) Drawing the direction by choosing another random live point. For
example, \citet{stokes2017new} uses a simplex-inspired walk to take
advantage of the distributions of live points. \citet{Pitkin2017}
combines of several proposals including uniform local step proposals,
differential evolution, and affine-invariant ensemble sampling MCMC
\citep{2010CAMCS...5...65G}. The affine-invariant ensemble sampler
is a popular choice for a MCMC proposal, because it does not need
tuning. This appears as a natural choice for NS, which already maintains
a population of points. However, in practice, this proposal performs
well primarily in Gaussian shapes, and \citet{Huijser2015} demonstrate
that in high dimensions, the sampler population collapses into a lower-dimensional
plane.

Many more geometric random walk algorithms exist and appear a-priori
suitable for LRPS in NS. For example, see the survey by \citet{Vempala05geometricrandom}.

\subsubsection{Number of steps for independence}

All aforementioned methods require the number of steps until a new,
supposedly independent point is found. How to chose the number of
steps? This is surely dependent on at least P/D issues. A simple technique
is to observe the change of $Z$ in a series of NS runs with increasing
number of steps \citep[e.g.,][]{Higson2019}.

Alternatively, that the live points are already uniformly distributed
suggests a simple tuning criterion. If the local MCMC chain has not
progressed further than the typical distance between two live points,
it has likely not stepped far enough \citep{Salomone2018}. A simple
auto-tuning method is thus to increase/decrease the number of steps
for the next iteration when that criterion is reached/not reached.
For the typical live point distance, \emph{UltraNest's} implementation
of HARM auto-tune method (``adapt=move-distance'') uses a mean Mahalanobis
distance of all pairs of live points. \citet{Salomone2018} emphasises
that for conserving detailed balance, a fresh run with no adaptation
has to be performed with a predefined number of steps, i.e., at least
as many as auto-tuning determined.

In CHMC and Galilean Monte Carlo, the reflections can lead to cycles
that do not explore more of the parameter space with increasing number
of steps. To address a similar problem in HMC, \citet{Hoffman2014NUTS}
proposed to construct forward and backward HMC trajectories only until
they turn back (No U-turn sampler, NUTS). Detailed balance is preserved
by randomly considering going forward or backwards while iteratively
doubling the number of steps. A U-turn can be identified when the
end point vectors show a positive dot product. A point is then sampled
from the trajectory, with acceptance probabilities determined by the
total energy (target probability and step momentum). Variants of NUTS
may be an interesting research direction for Demonic Nested Sampling
extensions.

To transfer this approach to NS, \citet{Griffiths2019} developed
the No Galilean U-Turn Sampler (NoGUTS). In the case of CHMC and Galilean
Monte Carlo under a flat prior, the momentum remains constant and
the target probability is either a constant or zero. Therefore, the
trajectories are straight until boundary reflections. This simplifies
the problem compared with NUTS. However, the constant shrinkage of
the sampling space in NS leads to biases if simple step size adaptations
are employed \citep{Griffiths2019}.

To allow good resolution of the sampling space and not stepping out
of the boundary too often, \citet{Skilling2012} recommends tuning
the step size so that most steps are accepted. This however means
that most of the steps constitute the progression of a straight line
and contain little additional information. A comparative study of
the mixing quality of NoGUTS, HARM and slice sampling in different
problems remains unstudied as of yet.

\subsection{Region sampling algorithms (non-MCMC)}

\label{subsec:Region-Sampling-Algorithms}

The safest way to sample from the prior under a likelihood constraint
is to sample randomly from the prior and reject the point if the likelihood
constraint is not fulfilled. This rejection sampling requires a good
proposal function to be efficient. In the case of LRPS, the support
of the proposal function corresponds to a parameter space region.
For correctness, it must fully contain the (unknown) likelihood contour,
i.e., the support of the likelihood-restricted prior. If a portion
of the parameter space is left out, as illustrated in Figure~\ref{fig:LRPSillustration},
the volume ratio $V_{i+1}/V_{i}$ will be overestimated. However,
the $Z$ estimate can be over or underestimated, depending on the
likelihood in the left-out region relative to the constructed region.
The methods presented in this section try to reconstruct this likelihood
contour itself, or at least a super-set of it.

\citet{Mukherjee2006} compute the smallest bounding ellipsoid that
contains all live points, and expands this by a factor ($\sim1.7$).
The left panel of Figure~\ref{fig:LRPSillustration} illustrates
such an ellipsoid wrapping the live points. Iteratively samples are
drawn from the ellipsoid and their likelihood is evaluated (rejection
sampling). This method thus has two parameters: The number of live
points, which helps trace out the parameter space, and the enlargement
factor. If the ellipsoid is not expanded enough, regions of the parameter
space are never sampled, if it is expanded too much, the rejection
sampling is inefficient. The choice of the enlargement is model and
dimension-dependent. Beyond approximating the minimum-volume bounding
ellipsoid with a scaled covariance metric, more advanced constrained
optimisation algorithms yield smaller volumes \citep{RollinsPliny2015}.
Why would ellipsoids be preferred over, say, a box \citep[used in][]{obrezanova2007gaussian,Moller2013}?
In the high-data regime, likelihood functions tend to become elliptical
distributions (such as a multi-variate Gaussian), which have ellipsoidal
contours \citep[see e.g.,][]{wilks1938large}.

If the problem presents multi-modality, the space between modes makes
the rejection sampling very inefficient. Therefore, \citet{Shaw2007}
cluster the live points with recursive k-means and employs ellipsoid
sampling for each cluster. In contrast, \citet{theisen2013analyse}
increases the number of clusters when the sampling efficiency drops
below a threshold. \citet{Feroz2008} further consider x-means, g-means
and pg-means for the clustering and uses x-means in the end. The algorithm
chooses the end points of the major axis of the ellipsoid enclosing
the live points, and attempts a k-means clustering with two clusters.
Live points are assigned to one of the two clusters, and used to construct
enclosing ellipsoids. If the two ellipsoids describe the live points
better than a single ellipsoid around all live points, the clustering
is accepted. This procedure is recursively repeated, until convergence.
The condition when a split is accepted needs to be defined, and a
variety of criteria are tested in \citet{Shaw2007,Feroz2008,Feroz2009},
including information criteria. The perhaps simplest is to consider
whether the volume is decreased by at least a certain factor, and
whether the ellipsoids are sufficiently apart. The MultiNest algorithm
has these criteria as parameters additional to the ellipsoid enlargement
factor. The left panel of Figure~\ref{fig:LRPSillustration} illustrates
the multiple-ellipsoids clustering.

\citet{Feroz2008} also considers adaptive enlargement factors that
decrease with iteration and ellipsoid volume, but these are ultimately
not used in the MultiNest algorithm presented in \citet{Feroz2009}.
The use of multiple ellipses is also useful for approximating peculiar
shapes. The efficiency has made MultiNest a popular algorithm, with
multiple implementations and interfaces, including PyMultiNest \citep{Buchner2014}
for Python, RMultiNest \citep{rmultinest} for R, an unnamed Mathematica
implementation \citep{Gervino2016}, DIAMONDS for C++ \citep{Corsaro2014},
JAXNS for the jax GPU programming language \citep{2020arXiv201215286A},
nestle \citep{nestle} and, derived from the latter, dynesty \citet{Speagle2020},
also for Python.

To avoid choosing an enlargement factor for every problem, \citet{Buchner2014stats}
estimates it from the live points by cross-validation. If some random
subset of live points were unknown, would we construct a region large
enough sample them? If not, the enlargement is insufficient and must
be increased. After several cross-validation rounds, a large-enough
enlargement is found. As a specific example, the RadFriends algorithm
\citet{Buchner2014stats} turns every live point into the centre of
an ellipsoid, whose shape determined by the choice of distance metric.
While the original RadFriends used spheres, \citet{Buchner2019c}'s
MLFriends uses the covariance of live points to define the ellipsoids,
leading to substantial speed improvements. The left panel of Figure~\ref{fig:LRPSillustration}
illustrates the region constructed by MLFriends. Clusters are naturally
defined by checking which live points are contained in other live
points' ellipsoids. Subtracting the cluster means from each live point,
and taking the resulting covariance improves the geometry further,
without requiring a dedicated clustering algorithm.

\subsection{Hybrid methods}

\label{subsec:Hybrid-methods}

Hybrid methods combine the region reconstruction with MCMC sampling.
These reduce the number of likelihood evaluations by excluding space
that is very likely outside the contour, while retaining the dimensionality
scaling of MCMC algorithms.

\citet{Feroz2008} combined MultiNest with the MCMC proposal of \citet{sivia2006data}
of 20 steps and tested problems in up to 100 dimensions. Interestingly,
they find that using a proposal tuned with local covariances is inferior.
Such hybrid combinations are available in dynesty \citep{Speagle2020}.

Different to MultiNest's x-means, \citet{Trassinelli2019} explores
Gaussian mixture method and the mean-shift clustering method with
MCMC, but lacks numerical comparisons to judge any improvements over
previous publications, both for the proposal and clustering method.

The \emph{UltraNest} package can combine two region construction methods,
MLFriends and a single ellipsoid. The former works well in low dimensions
and with multiple clusters, while the latter works well for Gaussian-like
likelihoods. The enlargement factors are estimated in both cases with
bootstrapping. New points are only considered when they fall within
both region constructions. Step sampling methods can then take advantage
of both regions to pre-filter proposed points to avoid model evaluations.

\section{Nested sampling variations}

\label{sec:Variations-Integration}

In standard NS, the sampler maintains a fixed number of live points.
This involves identifying the lowest likelihood point and replacing
it with LRPS, thereby increasing the likelihood threshold monotonically.
The following subsections look at variations of this scheme, including
soft likelihood constraints (§\ref{subsec:Softening-the-hard}), varying
the number of live points (§\ref{subsec:Vary-number-of}) and possibilities
for parallelisation (§\ref{sec:Parallelisation}) of the algorithm.

\subsection{Softening the hard likelihood constraint}

\label{subsec:Softening-the-hard}

LRPS methods deal with a hard likelihood threshold, and thus can only
test whether a point is acceptable or not. Points which turn out to
be below the contour are discarded, and it is difficult to know when
a random walk trajectory approaches the contour. At the same time,
when LRPS methods mistakenly exclude parameter space the shrinking
is accelerated, while slow LRPS sampling (such as too short random
walks) can cause points not to move enough, leading to slowed shrinkage.
To address these problems, methods have been developed which relax
the problem by avoiding a hard contour.

Importance sampling draws from an analytic shape to approximate the
unknown probability distribution of interest and reweighs the samples.
Importance Nested Sampling \citep[INS,][]{chopin2007contemplating,Chopin2008,Chopin2010}
applies the same concept by generalising nested sampling in this fashion.
\citet{Feroz2013} employed this concept with multi-ellipsoidal sampling,
and demonstrates that the use of otherwise discarded samples leads
to a substantially more precise $\log Z$ estimate. Indeed, the INS
estimator can somewhat correct for the incorrect LRPS sampling of
MultiNest in some difficult problems \citep{Buchner2014stats,Feroz2013}.
However, \citet{Nelson2020} demonstrate in an application to exoplanets
that both the standard NS and INS estimators in MultiNest can show
substantial scatter between runs beyond their uncertainties even in
the correct model (see their appendix A9, Figure 7 and 8). Indeed,
the scatter between runs can be an indicator that the LRPS is unreliable
(see §\ref{subsec:Inside-out}).

In Diffusive Nested Sampling \citep{brewer2011diffusive} the likelihood
contours created by shrinkages are reversibly explored. Particles
are not only allowed to traverse within the current likelihood constraint,
but to also to move up (down) in likelihood levels to a more (less)
constrained prior. The levels used are not based on all likelihoods
found. Instead a low number (dozens) of levels are maintained. Within
these levels, the sampled points are used to estimate the average
likelihood across the volume. This implies that relatively crude MCMC
proposals can be employed as LRPS procedures, as it is only required
that the level averages ultimately converge to the true value.

The up/down move of Diffusive Nested Sampling is achieved stochastically
with a Metropolis proposal. Thus Diffusive Nested Sampling turns the
entire nested sampling exploration into a MCMC process with arbitrary
precision. Having NS as a MCMC process makes it appealing for theoretical
(convergence) analysis. However, it also requires the use of MCMC
convergence diagnostics to determine the end point of a run, which
is not as well-defined as with standard nested sampling.

Demonic Nested Sampling \citep{Habeck2015demonic} softens the likelihood
contours by storing excess likelihood (energy) in a demon variable.
This can be combined with Hamiltonian Monte Carlo to store momenta,
and also delivers diagnostics about the state of the run through the
temperature evolution \citep[see also][]{Baldock2017}.

Beyond improving LRPS, the likelihood constraint may need to be softer
because the exact likelihood cannot be computed. For likelihood-free
inference, where only a stochastic but unbiased estimate of the likelihood
is available, \citet{Mikelson2020} present an NS variant.

\subsection{Varying the number of live points}

\label{subsec:Vary-number-of}

The original formulation of NS considered a population of live points
of fixed size. Upon finding disjoint clusters of live points, \citet{Feroz2008}
proposed to split the nested sampling run into separate, independent
runs. This procedure is adopted to prevent the loss of modes and increases
the number of live points to $N$ in each (potentially unequally sized)
mode. The volume associated with each sub-run needs to be estimated,
which is done numerically in MultiNest. However, although following
papers adopt the same procedure, the problem and prevention loss of
modes is not clearly demonstrated, and no test problem examples in
the literature are known to be prone to this issue.

A major slowdown of NS is that it needs to systematically progress
from the entire prior to the potentially extremely small posterior,
and this can take many iterations. This is in contrast to, for example,
MCMC, which can head quickly towards the posterior mass concentration,
or even be started in favourable locations. In some problems, especially
if one is confident only one posterior mode is present, it may therefore
be efficient to try to accelerate this ``finding'' phase in NS.

To address this, Dynamic Nested Sampling was proposed by \citet{Higson2017}.
First, NS is run with a fixed, low number of live points. Then, an
empirical CDF of the posterior weights is built, and an empirical
CDF indicating the fraction of the prior volume remaining above each
dead point. These two CDFs are linearly combined with ratios 1:3.
Then, the $10\%$ and $90\%$ quantiles, $L_{10\%}$ and $L_{90\%}$,
are sought which contain most of the CDF. A new NS run is started
from $L_{10\%}$, by creating $N'$ live points sampled by the LRPS
at that threshold. This NS run is then continued until $L_{90\%}$.
The new and original NS runs are merged, and the CDF procedure repeated
until some criterion is met, such as the $\log Z$ uncertainty. \citet{Higson2017}
demonstrate efficiency gains on toy and real-world problems. This
effectively employs an ad-hoc linear scalization to optimize a multi-objective
plan.

Considering the tree formulation presented in Section~\ref{subsec:NStree},
we can also interpret Dynamic Nested Sampling as repeatedly running
a tree search, and adding child nodes. The more general view is that
multiple agents could operate on the tree and until each of their
convergence criteria are met. We term this scheme Reactive Nested
Sampling, because an agent analyses the tree and reacts to its state.
For example, the Dynamic NS agent selects nodes just above $L_{10\%}$,
adds children to them and continues these new branches until the some
criterion is reached ($L>L_{90\%}$). In \emph{UltraNest}'s implementation,
three agents analyse the tree and react by adding children: Firstly,
the effective sample size criterion is improved by randomly sampling
nodes based on their posterior weights \citep{Higson2017}. Secondly,
the sampling uncertainty is improved by randomly sampling nodes based
on the information loss of leaving it out \citep[see][]{Speagle2020}.
Thirdly, improvements to the Z uncertainty are primarily limited by
phases with few live points, as the uncertainty on $\log Z$ is $\sigma=\sqrt{\sum_{i}\frac{1}{N_{i}^{2}}}$
\citep{Higson2017}. Therefore, the strategy identifies the minimum
$N$ required to reach the targeted $\log Z$ uncertainty, and enforces
that throughout the run. \citet{Speagle2020} analysed the uncertainty
estimations for different live point addition procedures.

As far as we can tell, there is no wrong way to insert new children,
so agents can aggressively optimize towards their criteria. Therefore,
computer science concepts such as intelligent agents and game theory
(e.g., minimax algorithms) can be considered. That said, running NS
initially with, to give an extreme example, $N=1$, is not advisable.
This is in part because multiple modes will not be explored, but also
because some LRPS procedures can substantially benefit from having
a few live points, so that they can build, for example, a covariance
matrix to estimate the current geometry or at least its rough scale.

\subsection{Parallelisation}

\label{sec:Parallelisation}

Taking advantage of multiple processing units can be achieved in several
levels: 

\paragraph{Parallelisation within the likelihood function}

The conceptionally simplest parallelisation is to use multiple computing
cores within the likelihood function, for example to process large
data sets or evaluate complex models. This requires no changes to
the basic NS algorithm. Related here is \citet{Graff2012}'s approach
of training a neural network to emulate the likelihood function during
a NS run to avoid evaluating the costly likelihood function when the
network accuracy is sufficient.

\paragraph{Parallelisation of the LRPS when its efficiency is low}

Inefficient searches for new live points can be parallelised by letting
multiple cores perform LRPS independently by worker processes \citep[e.g.,][]{Feroz2009}.
The first successful draw is accepted and returned to the main process.
Then the parallelisation is restarted with the subsequent threshold.
This method is easy to implement, as it requires little communication
and no modification of other NS components.

If an iteration simultaneously yields successful draws from multiple
workers, they can be accepted in order if they exceed the consecutive
thresholds. \emph{UltraNest} allows each processor to advance its
MCMC chain in parallel. When the required number of steps is reached
in one of the chains, the sample is accepted and the likelihood threshold
raised. The workers then find the last point in their chain which
fulfils the new likelihood threshold, and resume from there. This
avoids discarding the entire chain.

\paragraph{Parallelisation by adding and removing several live points at once}

This allows parallel searches for new samples \citep{burkoff2012exploring,Henderson2014,Martiniani2014}.
This is quantitatively tested for example in \citet{Baldock2017},
in addition to LRPS parallelisation. \citet{nielsen2013nested} presents
a modification of NS, where the number of removed live points is selected
to optimally divide the space based on the likelihood distribution
of live points.

\paragraph{Parallelisation by multiple independent runs with the same PRNG seed}

When attempting to draw a higher point, as many samples are drawn
from the constrained prior (via region sampling) as the number of
processes \citep{RollinsPliny2015}. Each process however only evaluates
the likelihood of the $p$-th sample, corresponding to its ID $p$.
Upon likelihood evaluation, each process distributes the result to
the others and decides locally for each sample whether to accept or
reject. Therefore, each process has the same nested sampling run but
the likelihood evaluations scale well with the number of processes.

\paragraph{Parallelisation with multiple independent runs}

Multiple, independent runs can be merged later \citep{skilling2009nested,henderson2017combined}.
Compared with the previous method, this has a higher number of effective
live points, and thus a higher accuracy. The drawback of independent
runs is that they cannot share information about the parameter space
geometry. The benefit of independent runs is that systematic errors
can be explored \citep{Higson2019}. \citet{Griffiths2019} take this
approach to the embarrassingly parallel extreme, with 20,000 independent
runs with only one live point each. They then identify runs that landed
in the same mode using a specialised algorithm, and unify those runs,
with the number of runs proportional to the mode probability. The
Diffusive Nested Sampling uses a similar parallelisation approach
by letting its walkers explore mostly independently \citep[B. Brewer, priv. comm]{brewer2011diffusive}.

\paragraph{Analysing similar datasets simultaneously}

\citet{Buchner2019c} noticed that when similar datasets are analysed,
their likelihood contours at a given iteration are similar. Computation
can be reduced by drawing in the joint likelihood contour. This is
most useful in large surveys of similar data or with Monte Carlo simulated
data under the same input parameters, when model evaluations are costly.
It could also be used to vary mildly the prior and likelihood assumptions
(T. Enßlin, priv. comm.), or the data preparation. 

Similarly, existing NS runs can be updated with a different likelihood
function, by reusing the successfully proposed live points until the
live point order diverges significantly. With similarity between live
point ordering measured by the normalized Kendall tau distance, such
a warm start is implemented in \emph{UltraNest}.

\section{Numerical experiment}

\label{sec:Numerical-experiment}

We briefly demonstrate numerically some aspects discussed. To that
end, we choose a toy problem with analytic integral, which demonstrates
some aspects of high-dimensional, highly informative problems. 

\begin{align*}
r_{1} & = & 10^{-11}\\
w_{1} & = & 0.4\times r_{1}\\
r_{2} & = & r_{1}/40\\
w_{2} & = & w/40\\
d_{1} & = & \sqrt{x^{2}+y^{2}}\\
d_{2} & = & \sqrt{\left(x+r_{1}\right)^{2}+y^{2}}\\
N_{i} & = & \frac{1}{\sqrt{2\pi w_{i}}}\exp\left[-\frac{1}{2}\left(\frac{d_{i}-r_{i}}{w_{i}}\right)^{2}\right]\\
L & = & N_{1}+100\times N_{2}
\end{align*}
\begin{figure}

\begin{centering}
\includegraphics[width=0.5\textwidth]{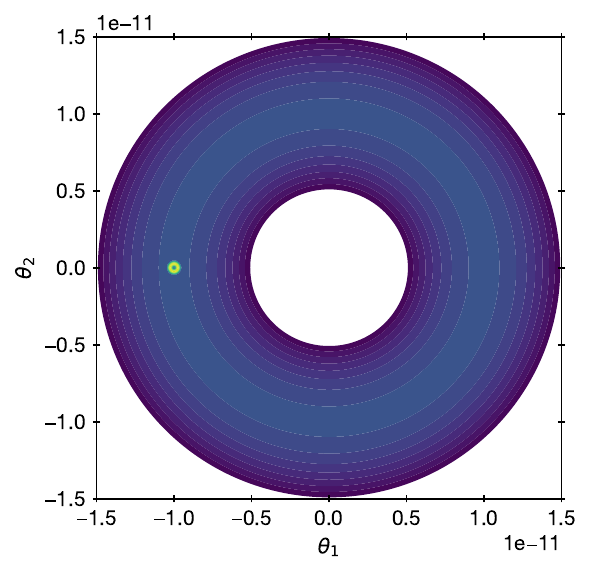}
\par\end{centering}
\caption{\label{fig:Toy-problem}Likelihood contours of the diamond ring toy
problem. A Gaussian shell (light green) is placed on a larger Gaussian
shell. This is an example of a small spike on a slab. Note the small
scales ($10^{-11}$) compared with the prior bounds $[-1,1]$.}

\end{figure}
The parameter space is two-dimensional, with $x$ and $y$ a priori
uniform between $-1$ and $+1$. The likelihood is visualised in Figure~\ref{fig:Toy-problem}.
The distance $d$ from the ring radius $r$ is compared with the width
$w$ using a Gaussian likelihood. The Gaussian shells make MCMC proposals
difficult, as they have to slowly wander along the shell. Two Gaussian
shells are added, with ratios $1:100$, making this a spike-and-slab
problem. The posterior is very narrow compared with the prior range,
requiring many NS iterations until the posterior is found.

\begin{figure}
\begin{centering}
\includegraphics[width=0.6\textwidth]{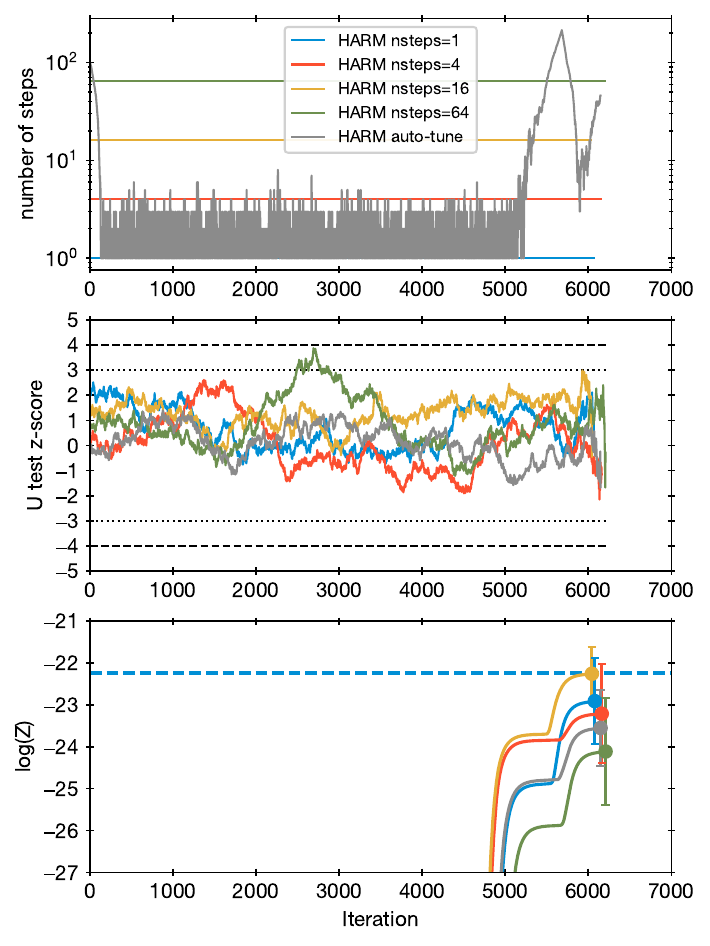}
\par\end{centering}
\caption{\label{fig:harm-steps}Number of MCMC steps. Five runs on the toy
problem are shown using HARM (Hit-and-Run Monte Carlo) method. \emph{Top
panel}: Number of steps per new sample. This is varied in the auto-tuning
method. \emph{Middle panel}: z-score of the U test computed in a rolling
window of 1000 iterations. \emph{Bottom panel}: Integral estimates.
Between iteration 5000 and 6000, a phase transition can be observed
where initially the integral estimate plateaus and then rapidly increases
again. The dashed line is the true value. }
\end{figure}

We run NS with a fixed, relatively low number of live points ($N=100$).
First, the number of MCMC steps in a LRPS method is tested. We choose
HARM, with the number of steps per NS iteration ranging from 1 to
64. We also add the auto-tune method presented in Section~\ref{subsec:Sampling-by-direction}.
Figure~\ref{fig:harm-steps} shows the run. In the top panel, the
auto-tuning method rises twice, namely when the two shells are encountered
and have to be navigated. Before, the number of steps could be tuned
to low values, allowing it to be orders of magnitudes more efficient
than using a safe number of steps throughout. The bottom panel of
Figure~\ref{fig:harm-steps} presents the accumulated integral estimate.
The final $\log Z$ estimate between the NS runs vary stochastically
in their final estimate, on a scale comparable to the uncertainty.
The error bars are relatively large because of the small $N$, but
are in acceptable ($<2\sigma$) agreement with the analytic value
(dashed horizontal line). The accumulated $\log Z$ shows a plateau
near iteration $5500$. This is the phase transition when the small
``diamond'' spike is being discovered on top of the larger ring
(slab). If runs had stopped before, based on a fixed computing budget
or because the live points appear similar, the additional probability
would not have been discovered. 

At the same time, the parameter space also becomes difficult to navigate
along the thin ring. The small euclidean distances traversed cause
the auto-tune method to increase its number of steps (grey curve rises
in the top panel). The U test is presented in the middle panel. The
curves vary mostly stochastically, not detecting strong deviations
even for the shortest MCMC step numbers. These curves look substantially
different, depending on what accumulation bandwidth is chosen (here:
1000 iterations), indicating that the insertion order does not see
problems with the sampling. This agrees with the mutually consistent
$\log Z$ estimates (error bars in bottom panel), which are usually
very sensitive to incorrect LRPS sampling.

Taking these results together, one may want to run HARM with at least
4 steps until the likelihood corresponding to iteration 5000, and
subsequently with 200 steps.

\begin{figure*}
\begin{centering}
\includegraphics[width=1\textwidth]{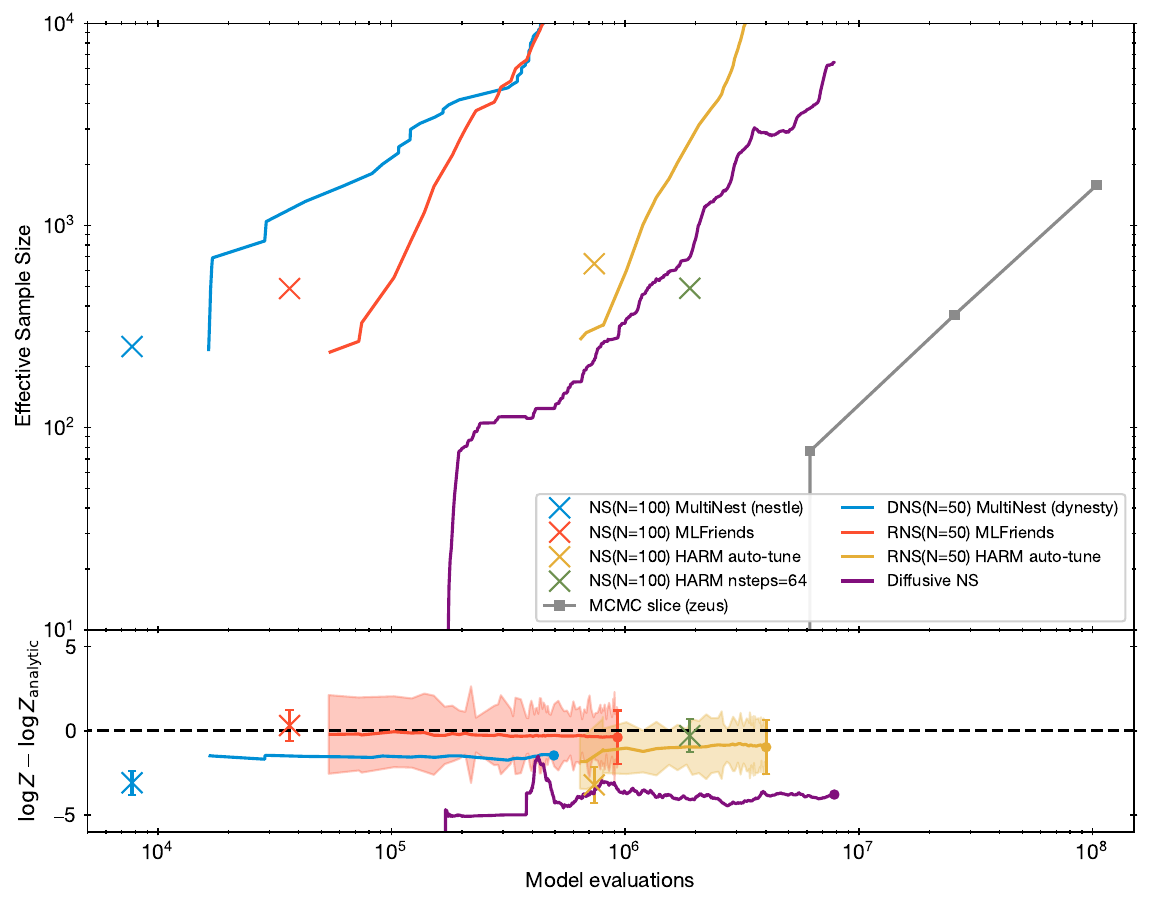}
\par\end{centering}
\caption{\label{fig:harm-cost}Efficiency comparison of different LRPS methods,
some using Dynamic NS or Reactive NS. \emph{Top panel}: Effective
number of samples extracted (upper left is better than lower right).
MCMC, after an initial phase of identifying the posterior, scales
approximately linearly. MLFriends (orange) and auto-tuning HARM (green)
show the steeper increase in samples per computation cost. \emph{Bottom
panel}: Integral estimate, relative to analytic value. The error bar
and shaded ares show $1\sigma$ uncertainty estimates, when available.}
\end{figure*}

The efficiency in extracting effective samples of various methods
is compared in Figure~\ref{fig:harm-cost}. To avoid giving a strong
geometric advantages to region-based methods, low $N$ are deliberately
chosen. MultiNest and MLFriends perform similarly within a factor
of three (blue, orange cross). As remarked above, HARM auto-tune (yellow
cross) is substantially more efficient than HARM with a large, fixed
number of steps per iterations (green cross).

With Dynamic Nested Sampling, started with $N=50$, and iteratively
adding $50$ live points, dynesty can efficiently extract more and
more effective samples (blue curve in Figure~\ref{fig:harm-cost}).
The Reactive NS methods developed here are used with an agent that
selects points based on posterior weights, and expands their parent
nodes. This is applied to MLFriends and HARM auto-tune (red and yellow
curves, respectively). Interestingly, the curve for MLFriends rises
super-linearly with likelihood cost. MLFriends scales in this problem
inversely with the number of live points. This is because the sampling
region can be approximated better, leading to smaller rejection rates.
Of course, this is only possible because of the low dimension of this
problem. MLFriends scales better than MultiNest here, because the
ellipsoid clustering is a poor description of the Gaussian shells.
There are of course problems where the converse is true, such as L-shaped
likelihood contours, where MultiNest substantially outperforms MLFriends.
Interestingly, also HARM auto-tune scales steeply with likelihood
evaluations. More live points mean the mean distance between them
is smaller, therefore the number of steps needed to reach a new point
is smaller. 

The Diffusive Nested Sampler DNest4 shows approximately a linear scaling
(purple curve in Figure~\ref{fig:harm-cost}). The number of likelihood
evaluations needed is higher than for the other methods. For comparison,
we also include a MCMC method, using the slice sampler zeus \citep{Karamanis2020}.
This shows a linear scaling (grey curve) after an initial phase where
the posterior needs to be identified. The efficiency of extracting
posterior samples in this case is substantially higher in NS methods,
because the MCMC auto-correlation is exceptionally poor. The computational
cost comparison in this problem is not representative of high-dimensional
problems, where the ranking of methods may be different and even reversed.

The bottom panel of Figure~\ref{fig:harm-cost} compares the integral
estimates. Here, the MultiNest estimate is off, as it excludes the
true value within its very small error bars. MLFriends and HARM methods
agree with the analytic value within $2\sigma$. The dynamic and reactive
nested sampling runs retain substantial uncertainties even with many
likelihood evaluations. This is because new live points are not added
to the initial phase of the run, causing the shrinkage estimates to
remain uncertain. DNest4 initially shows convergence to the true value,
but this trend is not improving later. DNest4 does not report uncertainties.

The MLFriends (for low-d) and HARM auto-tune (for low and high-d)
methods appear promising. They are implemented in the \emph{UltraNest}
Python package, with support for massively parallel computing on clusters.

\section{Summary}

This review has surveyed the nested sampling literature across many
subfields, and compiled proposed ideas and concepts. We have described
the problem types nested sampling is suitable for, and laid out the
practical difficulties implementations need to solve. To summarise:
\begin{enumerate}
\item Nested sampling is a practically useful algorithm for Bayesian model
comparison and parameter inference. It globally explores the parameter
space. This is important in problems with potentially complex, multi-modal
likelihoods. The exploration can largely proceed unsupervised and
without problem-specific tuning until a well-defined convergence point.
Nested sampling has some limitations to scale to very high dimensional
and highly informative problems.
\item This review highlighted the diversity of NS variants have been developed.
These include: NS without any MCMC (such as MultiNest and MLFriends),
using MCMC within NS, and running NS as a MCMC chain (Diffusive Nested
Sampling). A relatively recent development is to integrate HMC more
deeply, and adapting geometric random walk algorithms such as hit-and-run
and slice sampling.
\item Diagnostics and visualisations of the quality and correctness of runs
have improved substantially in the last few years.
\end{enumerate}

\section{Future research}

Much of the literature is involved with the application of nested
sampling to specific problems, evaluating its quality, and also proposing
new NS variants and implementations. Beyond this, we see the need
for systematic theoretical and practical evaluations in the future:
\begin{enumerate}
\item Investigating the theoretical foundations with an extended Sequential
Monte Carlo framework seems particularly promising. Given that NS
mutates only one particle, it should be investigated what theorems
can be transferred to such an extended framework.
\item Systematic numerical comparisons of a wide range of problems across
the PMDIT space are needed to judge the capabilities and limitations
of NS variants, including the LRPS methods and their tuning parameters.
\end{enumerate}
\addtocontents{toc}{\setcounter{tocdepth}{-1}}

\section*{Acknowledgements}

I thank the two referees, one of whom was Brendon Brewer, the associate
editor and the editor for their constructive comments which improved
the paper. I am very thankful to Josh Speagle for feedback and insightful
conversations. I am thankful to John Veitch, Matthew Griffiths, David
Wales for comments on the manuscript, and Michael Betancourt for insightful
conversations.

JB acknowledges support from the CONICYT-Chile grants Basal-CATA PFB-06/2007,
FONDECYT Postdoctorados 3160439 and the Ministry of Economy, Development,
and Tourism's Millennium Science Initiative through grant IC120009,
awarded to The Millennium Institute of Astrophysics, MAS. This research
was supported by the DFG cluster of excellence ``Origin and Structure
of the Universe''.

\appendix

\section{On encoding prior distributions}

\label{sec:Non-factorized-priors}

The unit hypercube transformation, $u\rightarrow\theta$, is a way
to encode priors by parametrising them in natural probability units,
$u_{i}\in[0,1]$. In factorised priors, the transformation is achieved
with inverse cumulative distribution functions, $\theta_{i}=F^{-1}(u_{i})$.
It is convenient as constant-energy trajectories are straight, and
sampling from geometrical shapes can be achieved without an additional
rejection. For this reason, some (but not all) popular NS implementations
let users specify priors via unit hypercube transformations, and in
many examples demonstrate only factorised priors. This has lead some
to believe that NS requires hypercubes and factorised priors. Here
we clarify that neither is the case.

Firstly, NS can proceed as long as a LRPS method and a likelihood
are defined. If the prior space is not $\mathbb{R}^{d}$, but arbitrary
``objects'', a simple example includes Metropolis samplers that
perturb the sample following the prior density. Indeed, such generic
treatments are possible in DNest4 \citep{Brewer2018DNest4}. This
includes allowing the dimensionality to vary, as demonstrated in \citet{Brewer2014}.

Secondly, we demonstrate two real-world cases where dependent priors
can be encoded in NS with unit hypercube transformations. Consider
a correlated Gaussian prior, defined by mean $\boldsymbol{\mu}$ and
covariance $\boldsymbol{\Sigma}$. The transform from a unit interval
$u\in[0,1]$ to an uncorrelated, standard normal prior can be encoded
through inverse cumulative distribution functions, $z_{i}=F^{-1}(u_{i})$.
Secondly, the transform from a standard normal prior to a general
Gaussian is performed with an affine transform, $\theta=\boldsymbol{A}z+\mu$,
where $\boldsymbol{\Sigma}=\boldsymbol{A}\boldsymbol{A}^{T}$. Therefore,
a correlated sample is obtained. This approach is not limited to multi-variate
Gaussians. For example, a Student-t would work the same and can receive
a degrees of freedom parameter. The generalisation of this approach
are copula models \citep[e.g.,][]{Nelsen2007}.

As a general approach, it can be useful to first transform one variable,
and then iteratively consider the conditional cumulative distribution
of the next given the previous. This iterative conditional approach
can also be applied if a non-analytic prior is only available as a
multi-dimensional histogram or samples $s$.

Often, procedures from random number generation can be adapted. As
an example, consider the problem of encoding fractions into the parametrisation.
This occurs for example when trying to fit for the elemental abundance
of some physical object, where the relative fractions must sum to
one. In some cases, it may be worthwhile to fit for an absolute parameter,
such as the total mass of that element, if that is closer to the observable,
and obtain fractions from the posterior. However, this requires placing
priors on the masses, which may be difficult. Lets therefore assume
we want to assume that the fractions each receive uniform priors,
but simultaneously must sum to unity. The appropriate distribution
for this scenario is the flat Dirichlet distribution with $\alpha=1$.
To obtain prior samples using a hypercube transformation, we obtain
independent gamma variables $z_{i}\sim\mathrm{Gamma}(\alpha,1)$,
i.e., $z_{i}=-\log u_{i}$, and obtain the fraction variables as $\theta_{i}=z_{i}/\sum_{i}z_{i}$.
More sophisticated transformations may be chosen to improve LRPS efficiency
\citep{Betancourt2012}.

We comment in passing that state-of-the-art HMC frameworks use similar
transformations internally to avoid trajectories exceeding the prior
support \citep{Carpenter2017}.

\section{Ellipsoidal sampling efficiency for ellipsoidal likelihoods}

\label{sec:Ellipsoidal-sampling-efficiency}

The computational cost of ellipsoidal nested sampling \citep{Mukherjee2006}
in the case of ellipsoidal likelihood contours is numerically explored,
following \citet{Allison2014}. Different to their fixed enlargement
treatment, here we determine the enlargement needed for rejection
sampling: We generate $N$ points from a hypersphere of dimension
$d$. In a bootstrapping scheme, we randomly leave points out \citep{Buchner2016}
and compute from the remainder a sample covariance matrix. This is
used to construct an ellipsoid, enlarged by the factor needed to recover
all left-out points. This is repeated 50 times, and the largest enlargement
factor stored. The ellipsoid volume ratio of the original sphere and
the constructed ellipsoid gives the acceptance rate of ellipsoidal
rejection sampling.

\begin{figure}
\includegraphics[width=0.8\textwidth]{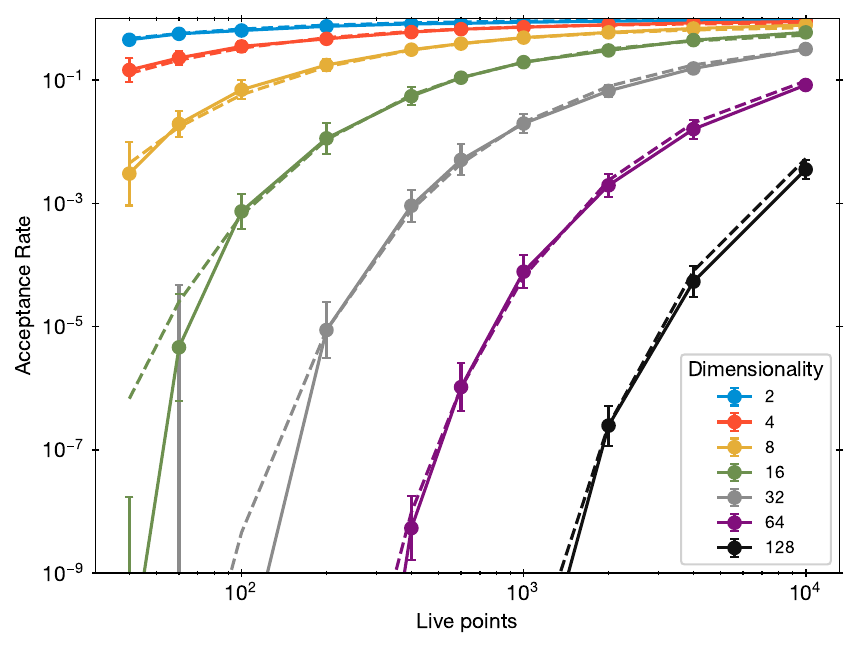}

\caption{\label{fig:ellipsoidalscale}Acceptance rate for ellipsoidal rejection
sampling. These measure the inverse excess volume of the constructed
ellipsoids to reliably sample the posterior. More live points lead
to tighter ellipsoids and higher acceptance rates. Increasing the
number of dimensions decreases the acceptance rate. The dashed curve
shows the empirical approximation of eq.~\ref{eq:empiricalscaling},
with the same colours as the data points for each dimensionality.}
\end{figure}

Figure~\ref{fig:ellipsoidalscale} presents the acceptance rate $\alpha$
as a function of $N$ and $d$. Error bars indicate the standard deviation
across 40 independently computed $\alpha$. Figure~\ref{fig:ellipsoidalscale}
shows that the acceptance rate increases with the number of live points
$N$, but decreases with dimensionality. Figure \ref{fig:ellipsoidalscale}
suggests that acceptance rates of $50\%$ can be maintained when $d$
doubles if $N$ increases five-fold. When $N$ is much smaller than
the dimensionality, this method breaks down dramatically. The data
points can be empirically described (dashed curves) by the following
formula:

\begin{equation}
\alpha=\left(1.07-\log d^{1/3}\right)\times\exp\left\{ -\left(\frac{6.83\times d^{1.9}}{N}\right)^{3/4}\right\} \label{eq:empiricalscaling-1}
\end{equation}
The first term decreases the acceptance rate only mildly with dimensionality,
and is therefore neglected in the cost scaling computations. The average
cost, i.e., the number of model evaluations needed, is $C_{\mathrm{ell}}=1/\alpha$.

\bibliographystyle{imsart-nameyear}
\bibliography{stats}

\end{document}